# Provisioning Energy-Efficiency and QoS for Multi-Carrier CoMP with Limited Feedback

Mohammad G. Khoshkholgh, Victor C. M. Leung, *Fellow, IEEE*, Kang G. Shin, *Life Fellow, IEEE*, Keivan Navaie, *Senior Member, IEEE*

## Abstract

We consider resource allocation (RA) in multi-carrier coordinated multi-point (CoMP) systems with limited feedback, in which a cluster of base stations (BSs) — each equipped with multiple antennas — are connect to each other and/or a central processor via backhauls/fronthauls. The main objective of coordinated RA is to select user equipments (UEs) on each subcarrier, dynamically decide upon the cluster size for each subcarrier, and finally partition the feedback resources — provisioned for acquisition of channel direction information (CDI) across all subcarriers, active cells, and selected UEs — in order to maximize the *weighted sum utility* (WSU). We show how to recast the WSU maximization problem to achieve spectral efficiency, quality-of-service (QoS), and energy-efficiency (EE). Specifically, we investigate four instances of WSU to maximize practical system objectives: (*i*) weighted sum capacity, (*ii*) weighted sum effective capacity, (*iii*) weighted sum energy-efficiency (EE), and (*iv*) weighted sum effective EE. The unified composition of these problems through WSU allows us to use the same set of developed algorithms for all cases. The algorithms have a greedy structure achieving fast convergence, and successfully cope with the huge computational complexity of RA problems, mostly rooted in their combinatorial compositions. Our simulation results shed lights on the network optimization by discovering insights on appropriate cluster-size, distribution of BSs in the cluster, and the number of subcarriers. The proposed UE scheduling and subcarrier assignment are shown to improve the system performance by several orders-of-magnitude.

## I. INTRODUCTION

Fifth-generation (5G) cellular systems are expected to witness a 1000x capacity increase over 4G's performance in order to handle the forecasted traffic demands by 2020 [1]. Hundreds of exabytes IP traffic will not be solely a ramification of the ever-growing demands for high-definition video streaming, social networking, real-time interactions, and the like, but partly induced by the materialization of prevalent services such as Internet of Things (IoT) [2], industrial Internet, and ubiquitous machine-to-machine (M2M) communications [3]. To achieve such improvement in capacity, in addition to multitude upgrades, 4G standards undergo a number of architectural and component revolutions [1, 4]. Disruptive communication technologies in achieving this hefty goal include ultra-dense heterogenous cellular networks (HCN) [1, 5], millimeter wave communications [6], massive multi-input multi-output (MIMO) communications [7], coordinated multi-point (CoMP) communications, *a.k.a.* networked/virtual MIMO, [8, 9]. From networking perspectives, mobile-edge computing, software-defined networking, and network function virtualization are considered as game changing innovations [1, 10].

While capacity, latency, reliability, and mobility are the principal drivers for these disruptive innovations [10], it is yet to fully contemplate how and to what extent 5G standards will meet other critical concerns, such as *energy-efficiency* (EE). Intuitively, to keep the transmission power consumption at the current levels, 5G standards should augment the corresponding EE up to 1000x [1]. Considering the inherent trade-offs between capacity and EE, this ambitious goal is extremely challenging—albeit not impossible—to achieve [11, 12]. Consequently, investigation of the EE in 5G is necessary, thus has triggered a broad range of research activities [11, 13]. For instance, a multi-objective approach has been considered in [11] for comprehensive investigation of the trade-off between EE and spectral efficiency. To reduce transmission energy of HCNs, the authors of [13, 14] proposed to offload traffic between macro and micro BSs. In [15], infrastructure-on-demand is introduced to maximize the energy-savings at BSs and other infrastructure components.

In concert with this efforts, in this paper, we study EE in 5G[1]. In particular, we focus on CoMP systems while effectually coordinating the adverse effects of extreme inter-cell interference caused by aggressive frequency reuse [16–18]. In practice, some may, however, argue against the claimed capacity of CoMP systems [17, 19]. The substantial difference of spectral efficiency between analysis and measurement is perhaps because the coordinated techniques were not entirely a native design premise of 4G standards, but they are fundamental design postulates for 5G networks. On the other hand, their importance is gaining momentums as they are the key enablers of emerging technologies, such as centralized- and cloud-radio access networks (C-RANs).

### A. Related Work

The literature of CoMP communications is rich and promising, e.g, [20–22]. In [21] a novel dynamic clustering and interference coordinator system, called NEMOx, is proposed and respectively prototyped. Energy-driven resource allocation (RA) and beamforming in CoMP systems has also been studied extensively [23–25]. Energy-efficient RA problems are typically non-convex, thus they are often very challenging. In [23, 25] the EE RA problem is transformed into a subtractive form which is more amenable for analytical investigation and requires manageable computational complexity. The authors of [23] also observ that for some particular scenarios, an EE solution may also lead to a spectrum-efficient solution.

Dynamic clustering has investigated further in [26–29]. In fact, dynamic clustering, in conjunction with UE scheduling and beamforming, is a crucial enabler for no-edge cellular systems. Dynamic clusters however are very challenging to be modelled, and often incorporate very high computational complexity, see, e.g., [26–28]. For cluster management, the authors in [26] recommended composition of clusters out of a given set of candidate BSs. In practice, besides the complexity issues, limiting factors are large volume of traffic needs to be shared on the backhauls, and the required rigid synchronization among the BSs. Note that backhaul energy consumptions due to shared data traffic might generate an energy bottleneck in the network [30, 31]. One solution to this challenge is to switch off as many BSs as possible without sacrificing signal-to-interference plus noise ratio (SINR) requirements of UEs, as shown in [32, 33]. To manage the energy consumption, [33] proposes minimizing the transmission power subject to UE's SINR requirements. Greedy port selection jointly with group sparse beamforming then proposed. Apart from extremely high computational complexity of such approaches especially in crowded scenarios, they often fails to achieve the advantages of ultra-dense networking by turning off many BSs and adjusting the operating points of beamforming or power control to the minimum required SINR thresholds, which is not in full compliance with the huge capacity requirement of 5G networks. Moreover, there is no guarantee that minimizing the BS's power transmission—commonly considered as the main objectives in these lines of research—will lead to the minimization of backhaul's energy consumption or the maximization of the network's EE. To tackle this, some researchers, on the other hand, advocat coordinated schemes that entirely rely upon the usage of backhauls for signaling and channel direction information (CDI) sharing [23, 34, 35]. In this paper, we consider this recommendation in devising coordinated RA solutions.

---

[1]Note that we do not explicitly incorporate backhauls' energy consumption in our analysis, however the proposed system model is backhaul-friendly.

In the above-mentioned approaches, availability of accurate CDI at the BSs/central processor often plays an important role. This is often impractical especially in frequency division duplex (FDD) systems, see, e.g., [36–40]. In practice, UEs need to quantize the CDIs before transmission through designated feedback channel with limited capacity resources. Optimizing the required feedback resources in a two-cell coordinated setting is investigated in [40], where the objective is to maximize the expected SINR. In [41], the notion of random clustering in limited-feedback CoMP systems is proposed. Proper feedback bit partitioning (FBP) is shown to be able to reduce the capacity gap—the difference between the capacity of full CDI and that of with quantized CDI—caused by CDI quantization errors and delay. Nevertheless, BSs might have different versions of the acquired quantized CDIs due to communication impairments imposed by backhauls. This issue is investigated in [42] where theoretical bounds derived indicating the growth of the slope of the capacity gap with respect to the transmission power.

## B. Main Contributions

In the related literature, energy-driven RA in multi-carrier limited-feedback CoMP (McLf-CoMP) systems encompassing traits of QoS, despite its growing importance, has not been addressed well. For the case of maximizing the capacity (or capacity gap) it is commonly presumed that subcarrier assignment (SA) and UE scheduling is pre-specified/given [34, 41, 43, 44], which is not the case of reality. Some may advocate UE scheduling via adopting the methods originally coined by authors in [45]. Such methods, nevertheless, require the availability of quantized CDIs from the whole UEs that are apparently not scalable, particularly in populated multi-cell scenarios with limited-feedback. Besides, the incorporation of dynamic cluster-size in developing FBP mechanisms is commonly overlooked, e.g., it is often assumed that the cluster-size is fixed [34, 43, 44] or random [41]. In reality, SA, UE scheduling, size of the cluster, and FBP are interacting with each other and should be effectually integrated in RA policy.

Note that in many of the above mentioned work, for simplicity, it is mainly assumed that the feedback resources allocated to each UE is known/pre-specified, see, e.g., [34, 41, 43, 44, 46]. On account of limited feedback resource per cluster[2] in this paper we propose treating the FBP along with coordinated allocating of the other resources such as subcarriers and BSs' beams.

CDI quantization and its impact on the capacity gap is commonly investigated in the literature, see, e.g., [34, 37–41, 43, 44, 46]. Nevertheless, it is not straightforward to extend the results to the advanced cases where the performance measures is e.g., effective capacity—which addresses the statistical QoS requirement [47]—or EE. In this paper, we further investigate CDI quantization and its impact on the system effective capacity and EE.

For easy reference our main contributions are summarized as follows.

- Formulation of *weighted sum utility* (WSU) problem as an analytical framework to model complex system performance objectives. We then demonstrate how this general analytical framework is capable of incorporating crucial performance objective in the RA problems, including: (*i*) weighted sum capacity (WSC), (*ii*) weighted sum effective capacity (WSEC), (*iii*) weighted sum energy-efficiency (WSEE), and (*iv*) weighted sum effective energy-efficiency (WSEEE). A unified optimization setting allows us to utilize the same set of RA algorithms.

- Derivation of computationally affordable yet sufficiently accurate bounds on the capacity, and effective capacity in the McLf-CoMP systems.

- Construction of several greedy algorithms called *greedy-FBP* (gFBP)—for partitioning feedback bits among subcarriers, cells, and UEs—and Cluster-based SA (C-bSA)—for per-subcarrier cluster size determination (p-sCSD), UE scheduling, and SA. These algorithms are able to effectively regulate high computational complexity of the RA problems, rooted in their combinatorial structures as well as interactions among SA, UE selection, p-sCSD, and FBP as well as the sheer size of the problems.

- Detailed simulation to shed lights on the significant benefits of jointly scheduling UEs and p-sCSD. Insights on the position of BSs and the size of the cluster are further demonstrated.

---

[2]For instance, in the case of distributed antenna systems (DAS) and C-RAN the BSs belong to the cluster usually have the same cell ID. In such cases feedback resources are required to be managed per cluster.

## C. Organization of the Paper

The rest of this paper is organized as the following. In Section II we present the system model. Problem formulation is presented in Section III. In Section IV we then look at resource allocation problem with various objectives and for a unified resource allocation problem. Algorithms are presented in Section V to find the solutions to the resource allocation problem. Numerical and simulation results are presented in Section VI followed by conclusions in Section VII.

## II. SYSTEM MODEL

### A. Network Model

In this paper we study resource allocation (RA) in coordinated multi-cell OFDMA systems while focusing on the downlink communication. The bandwidth is partitioned into $|\mathcal{N}|$ statistically independent subcarriers indexed by $n \in \mathcal{N}$. We consider block fading and assume that the fading on each subcarrier $n$ is flat [48, 49].

*Remark 1:* Note that we consider orthogonal frequency division multiplexing (OFDM) for the modulation part and OFDMA for the access part. This choice may not be the best as there are already other tentative modulation schemes with higher spectral efficiency and/or lower peak-to-average ratio, such as filter band multi-carrier, that may eventually become better options for 5G systems [1, 50].

A cluster consists of $|\mathcal{C}|$ heterogenous base stations (BSs), indexed by $c \in \mathcal{C}$. BSs are connected to a central processor via high-speed, low latency backhauls/fronhauls (e.g., microwave or fiber-optic connections). Each BS is equipped with $N_t$ transmit antennas. For CoMP to be feasible, we assume $N_t > |\mathcal{C}|$ [34, 38]. There are $|\mathcal{U}|$ single-antenna EUs indexed by $u \in \mathcal{U} = \{1, 2, \dots, |\mathcal{U}|\}$, randomly scattered in the cluster's coverage area. The UEs are then associated with appropriate BSs according to an existing cell association mechanism—which is not the focus of this paper—to partition set $\mathcal{U}$ into $|\mathcal{C}|$ disjoint sets $\{\mathcal{U}_c\}$. Let $\mathcal{U}_c$ be the UEs served by BS $c$. We also assume the communication system excludes joint transmissions, i.e., $\mathcal{U}_c \bigcap \mathcal{U}_{c'} = \emptyset$ . As a result, each UE $u_c \in \mathcal{U}_c$ is solely served by BS $c$ while the rest of the BSs set their beam patterns to pre-cancel the interference on UE $u_c$, which is refereed to as *coordinated beamforming* [20]. Although its spectral efficiency is lower than the joint transmission, it has been recommended repeatedly in the literature, e.g., [23, 34, 35, 38], as it reduces loads on the backhauls and alleviates the need for tight synchronization among the BSs. The distance-dependent path-loss attenuation between UE $u_c$ and BS $c$, $\rho_{u_c c}$, is proportional to $(1 + d_{u_c c})^{-\nu_{cc}}$ where $\nu_{cc} \in (2, 6)$ is the path-loss exponent associated with BS $c$ and cell $c$, and $d_{u_c c}$ is the Euclidean distance between the transceiver pairs. Likewise, the path-loss attenuation between interfering BS $c'$ and UE $u_c$ is expressed as $\tilde{\rho}_{u_c c'} = (1 + d_{u_c c'})^{-\nu_{cc'}}$ where $\nu_{cc'} \in (2, 6)$. On each sub-channel $n$, the transmitted signal undergoes flat-fading fluctuation denoted as $\boldsymbol{h}_{u_c c, n} \in \mathbb{C}^{N_t \times 1}$ with the CDI (direction) $\tilde{\boldsymbol{h}}_{u_c c, n} = \boldsymbol{h}_{u_c c, n} / \|\boldsymbol{h}_{u_c c, n}\|$ ($\|.\|$ denotes norm-2). On this subcarrier, the interfering links are denoted by $\boldsymbol{g}_{u_c c', n} \in \mathbb{C}^{N_t \times 1}$ where $c' \neq c$. We assume $\boldsymbol{h}_{u_c c, n}$ and $\boldsymbol{g}_{u_c c', n}$ are independent, and their elements are complex Gaussian random variables with zero mean and variance of 1, which are Rayleigh assumptions [34, 37, 38]. Moreover, subcarriers are statistically independent, and free of inter-symbol interference (ISI) as well as inter-carrier interference (ICI) [48].

## B. Quantizing the Channel Direction Information and Coordinated Beamfoming

We focus on frequency division duplex (FDD), e.g., LTE-Advance. For such systems, the acquisition of channel direction information (CDI) often requires a feedback channel in the uplink. Let $B_{tot}$ be the total assigned feedback capacity (in number of bits) to the cluster [36].

Consider subcarrier $n$ and, for the time being, assume a specific partitioning of the feedback bits.

At the start of each transmission slot BSs sequentially broadcast pilot signals to facilitate CDI estimation at the UEs. Devoting adequate pilot resources, UEs are able to estimate the corresponding CDIs with acceptable accuracy.

UE $u_c$ quantizes CDIs according to previously constructed quantization code-books for intended CDIs, $\mathcal{W}_{u_c c, n}$, and interfering CDIs, $\{\mathcal{W}_{u_c c', n}, \forall c' \neq c\}$. UEs then transmit the indices of the selected code-words to the corresponding BSs on the designated feedback channel. Similar to [38] here, CDIs are quantized separately at each UE. Let $\mathcal{C}_n^a \subseteq \mathcal{C}$ contain the index of active BSs on subcarrier $n$. Each UE, $u_c$, requires $\sum_{c' \in \mathcal{C}_n^a} B_{u_c c', n}$ feedback bits to quantize the CDIs $\tilde{\boldsymbol{h}}_{u_c c, n}, \tilde{\boldsymbol{g}}_{u_c c', n} \ \forall c' \in C_n^a / \{c\}$ and report them to BS $c$. We denote $\hat{\boldsymbol{h}}_{u_c c, n} = \max_{l=1,\dots,2^{B_{u_c c, n}}} |\tilde{\boldsymbol{h}}_{u_c c, n}^\dagger [\mathcal{W}_{u_c c, n}]_{(:, l)}|^2$, where $[\mathcal{W}_{u_c c, n}]_{(:, l)}$ refers to the $l$-th column of code-book $\mathcal{W}_{u_c c, n}$, i.e., the $l$-th code-word, and superscript $\dagger$ indicates transpose conjugate as the quantized CDI of $\tilde{\boldsymbol{h}}_{u_c c, n}$. Similarly the quantized CDI of $\tilde{\boldsymbol{g}}_{u_c c', n}$ is defined as $\hat{\boldsymbol{g}}_{u_c c', n} = \max_{l=1,\dots,2^{B_{u_c c', n}}} |\tilde{\boldsymbol{g}}_{u_c c', n}^\dagger [\mathcal{W}_{u_c c', n}]_{(:, l)}|^2$.

If $u_c$ is selected by the central processor for the transmission, it then reports the indices $l_{u_c c, n}^* = \arg\max_l |\tilde{\boldsymbol{h}}_{u_c c, n}^\dagger [\mathcal{W}_{u_c c, n}]_{(:, l)}|^2$, and $\{l_{u_c c', n}^* = \arg\max_l |\tilde{\boldsymbol{g}}_{u_c c', n}^\dagger [\mathcal{W}_{u_c c', n}]_{(:, l)}|^2\}$ to the BS $c$. BS $c$ then de-quantizes $l_{u_c c, n}^*$ to $\hat{\boldsymbol{h}}_{u_c c, n}$ and $\{l_{u_c c', n}^*\}$ to $\{\hat{\boldsymbol{g}}_{u_c c', n}\}$, and shares $\{\hat{\boldsymbol{g}}_{u_c c', n}\}$ with the other BSs.

Other BSs also provide BS $c$ with the interfering channel directions between BS $c$ and selected UEs in other cells, $\hat{\boldsymbol{g}}_{u_{c'} c, n} \ \forall u_{c'} \notin \mathcal{U}_c$. Upon receiving required CDIs, BS $c$ constructs matrix $\hat{\boldsymbol{G}}_{u_c c, n} = [\hat{\boldsymbol{g}}_{u_{c'} c, n} \forall u_{c'} \notin$ $\mathcal{U}_c] \in \mathbb{C}^{N_t \times (|\mathcal{C}_n^u|-1)}$ and produces coordinated beamforming vector $\hat{\boldsymbol{f}}_{c,n} \in \mathbb{C}^{N_t \times 1}$ as [34]

$$\hat{\boldsymbol{f}}_{c,n} = \frac{(\boldsymbol{I} - \mathcal{P}(\hat{\boldsymbol{G}}_{u_c c,n}))\hat{\boldsymbol{h}}_{u_c c,n}}{\|(\boldsymbol{I} - \mathcal{P}(\hat{\boldsymbol{G}}_{u_c c,n}))\hat{\boldsymbol{h}}_{u_c c,n}\|}, \tag{1}$$

where $\mathcal{P}(\hat{\boldsymbol{G}}_{u_c c,n}) = \hat{\boldsymbol{G}}_{u_c c,n}(\hat{\boldsymbol{G}}_{u_c c,n}^{\dagger}\hat{\boldsymbol{G}}_{u_c c,n})^{-1}\hat{\boldsymbol{G}}_{u_c c,n}^{\dagger}$ is the projection operator. This beamforming technique is commonly referred as *inter-cell interference cancellation* (ICIC) [34, 38]. ICIC is in fact an extension of zero-forcing beamforming (ZFBF) for the coordinated systems. Note that the above model assumes that BSs possess an extent of processing capabilities and capable of obtaining beamforming vectors. In the case of distributed antenna systems (DAS) the central unit obtains the beamforming vectors.

In this model, SINR at $u_c$ is

$$\gamma_{u_c c,n} = \frac{\rho_{u_c c}\frac{P_{c,n}}{N_t}|\boldsymbol{h}_{u_c c,n}^{\dagger}\hat{\boldsymbol{f}}_{c,n}|^2}{\sigma^2 + \sum\limits_{c' \in \mathcal{C}_n^c/\{c\}} \bar{\rho}_{u_c c'}\frac{P_{c',n}}{N_t}|\boldsymbol{g}_{u_c c',n}^{\dagger}\hat{\boldsymbol{f}}_{c',n}|^2}, \tag{2}$$

where $\hat{\boldsymbol{f}}_{c',n}$ is the coordinated beamforming vector constructed at BS $c'$ and is independent of $\hat{\boldsymbol{f}}_{c,n}$, $\sigma^2$ is AWGN and inter-cluster interference contribution, and $P_{c,n}$ is the transmission power at BS $c$ on subcarrier $n$. Since BSs do not access the channel quality information (CQI), the allocated power on each subcarrier is simply divided by the number of antennas, $N_t$. Furthermore, due to mismatch between $\hat{\boldsymbol{g}}_{u_c c',n}$ and $\boldsymbol{g}_{u_c c',n}$, there holds $|\boldsymbol{g}_{u_c c',n}^{\dagger}\hat{\boldsymbol{f}}_{c',n}|^2 \neq 0$ while $|\hat{\boldsymbol{g}}_{u_c c',n}^{\dagger}\hat{\boldsymbol{f}}_{c',n}|^2 = 0$. To model this mismatch and incorporate its impact in the system design, here we adopt Quantization Cell Approximation (QCA) [38, 45, 51]. This implies that for the available feedback bits $\{B_{u_c c',n}, \forall c' \in \mathcal{C}_n^a\}$ on subcarrier $n$ the quantization error $\sin^2(\theta_{u_c c,n}) = \min_l |\tilde{\boldsymbol{h}}_{u_c c,n}^{\dagger}[\mathcal{W}_{u_c c,n}]_{(:,l)}|^2$ has the following probability density function (pdf) [45]

$$f_{\sin^2(\theta_{u_c c,n})}(x) = (N_t - 1)2^{-B_{u_c c,n}}x^{N_t-2}, \ \ 0 \leq x \leq \delta_{u_c c,n}, \tag{3}$$

where $\delta_{u_c c,n} = 2^{-\frac{B_{u_c c,n}}{N_t - 1}}$. The same is similarly obtained for pdf of $\sin^2(\theta_{u_c c',n}) = \min_l |\tilde{\boldsymbol{g}}_{u_c c',n}^{\dagger}[\mathcal{W}_{u_c c',n}]_{(:,l)}|^2$.

## III. Problem Formulation

The radio resource allocation (RA) problem is formulated to optimize an objective function by allocating the subcarriers, size of the cluster, and the feedback capacity in McLf-CoMP systems. We first elaborate on the system constraints in the optimal RA problem.

According to the OFDMA technique, at most one UE is scheduled in each cell $c$ and on each subcarrier $n$. Intuitively, to exploit the frequency, time and multi-user diversities inherent in multi-user OFDMA systems, subcarriers should be assigned by accounting for the impact of the performance of communication links on the overall network performance.

Besides, the performance of a communication link in McLf-CoMP systems is greatly influenced by feedback bit partitioning (FBP) and per-subcarrier cluster-size determination (p-sCSD). It is partly because UE selection in a cell is interrelated with the same procedure in other cells, i.e., a portion of each BS's degrees-of-freedom is actually devoted to pre-canceling intra-cell interference. Determining the cluster size on each subcarrier is also affected by the cluster size on the other subcarriers as well as FBP for two reasons. First, by switching off some of the BSs in the cluster, the selected UEs need not quantize/report the corresponding CDIs, thus saving the scarce feedback resources which could be used to enhance the resolution of CDI quantization of active BSs. Second, intelligently turning off the BSs reduces inter-cell interference. Nevertheless, by reducing the number of active BSs on a subcarrier, there is a chance of decreasing the net performance since some UEs ended up being disconnected form the network. It is, therefore, necessary to handle FBP, SA, and p-sCSD by carefully considering the trade-offs between the merits and costs.

Let $x_{u_c c, n} \in \{1, 0\}, c \in \mathcal{C}, u_c \in \mathcal{U}_c$ denote the SA indicator on subcarrier $n$. Further, let $y_{c,n} \in \{0, 1\}$ denote the indicator determining the on/off status of BS $c$ on subcarrier $n$. We also define set $\mathcal{C}_n^a = \{c \in \mathcal{C} : y_{c,n} = 1\}$ as the active BSs on subcarrier $n$. SA and p-sCSD constraints are, respectively, prescribed as

$$C1 : y_{c,n} \sum_{u_c \in \mathcal{U}_c} x_{u_c c, n} \leq 1, \forall n \in \mathcal{N}, c \in \mathcal{C}, \tag{4}$$

$$C2 : \sum_{c \in \mathcal{C}} y_{c,n} \leq |\mathcal{C}|, \forall n \in \mathcal{N}. \tag{5}$$

Note that if $\sum_{u_c \in \mathcal{U}_c} x_{u_c c, n} = 0$, then $y_{c,n} = 0$. On the other hand, if $y_{c,n} = 0$, then $x_{u_c c, n} = 0 \ \forall u_c \in \mathcal{U}_c$ on subcarrier $n$. Additionally, denote $B_{u_c c', n} \in \{0, 1, 2, \ldots, B_{tot}\} \ \forall c' \in \mathcal{C}_n^a$ is the assigned number of feedback-bits to the UE $u_c \in \mathcal{U}_c$ on subcarrier $n$, where $B_{tot}$ as the total available feedback capacity designated to the cluster, we then have

$$C3 : \sum_{n \in \mathcal{N}} \sum_{c \in \mathcal{C}} y_{c,n} \sum_{u \in \mathcal{U}_c} x_{u_c c, n} \left( \sum_{c' \in \mathcal{C}_n^a} B_{u_c c', n} \right) = B_{tot}. \tag{6}$$

We refer to (6) as the FBP constraint. Note that the total assigned feedback bits to UE $u_c \in \mathcal{U}_c$ on subcarrier $n$ that is $\sum_{c' \in \mathcal{C}_n^a} B_{u_c c', n}$ should be optimally partitioned among the BSs in $\mathcal{C}_n^a$. Such partitioning, on the other hand, determines the size of the quantization code-books at UE $u_c$ and in turn the accuracy of the CDI quantization procedure. Referring to (6), we recognize the effect of p-sCSD, UE scheduling and SA on the FBP constraint.

*Remark 2:* Note that in the literature, see e.g., [34, 37–41, 43, 44, 46], it is often assumed that the feedback resource is given per each UE, i.e., $\sum_{c' \in \mathcal{C}_n^a} B_{u_c c', n} = \bar{B}$ where $\bar{B}$ is pre-specified threshold. In contrast, according to constraint (6) it is then the RA's responsibility to dynamically derive how many feedback bits are available per UE. More importantly, FBP constraint implies that the available feedback bits should be partitioned coordinately among the cells. Besides relevancy to the important practical scenarios such as DAS and C-RAN systems, FBP constraint can provide much higher system performance compared to the cases in which BSs belong to the same cluster receive pre-specified share of feedback resource.

Let $X^{|\mathcal{U}| \times |\mathcal{N}|} = [x_{u_c,n}]$ be the SA matrix, $Y^{|\mathcal{C}| \times |\mathcal{N}|} = [y_{c,n}]$ be the BS activity matrix, and for each subcarrier $n$ $B_n^{|\mathcal{U}| \times |\mathcal{C}|} = [B_{u_c c,n}]$ be the FBP matrix. Let $\Pi$ denote the set in which all constraints C1–C3 are satisfied. Having the resource allocation constraints specified, we are now ready to introduce the central objectives of the RA problems as follows.

Let $\Lambda_{u_c,n}(\{X, Y, \{B_n\})$ denote the utilization corresponding to UE $u_c$ on subcarrier $n$. By introducing weights $w_c$, the following optimization problem weighted sum utility (WSU) is the general RA for the system considered in this paper:

$$\text{WSU}: \max_{\{X, Y, \{B_n\}\} \in \Pi} \sum_{c \in \mathcal{C}} \omega_c \sum_{n \in \mathcal{N}} y_{c,n} \sum_{u_c \in \mathcal{U}_c} x_{u_c c,n} \Lambda_{u_c,n}(\{X, Y, \{B_n\}). \tag{7}$$

We will henceforth drop parameters in $\Lambda_{u_c,n}(\{X, Y, \{B_n\})$ and the counterpart performance measures. We further define $\lambda_n$ as $\lambda_n = \sum_{c,u_c} \omega_c y_{c,n} x_{u_c c,n} \Lambda_{u_c,n}$ as the network utilization on subcarrier $n$, and $\lambda = \sum_n \lambda_n$ as the total network utilization.

## IV. Resource Allocation Problems

We consider the following four network utility functions: (*i*) weighted sum capacity (WSC); (*ii*) weighted sum effective capacity (WSEC); (*iii*) weighted sum energy efficiency (WSEE); and (*iv*) weighted sum effective energy efficiency (WSEEE). Next we will investigate the RA problems corresponding to each of these objectives.

*1) Weighted Sum Capacity (WSC):* Let $\bar{R}_{u_c c,n}$ be the capacity of UE $u_c \in \mathcal{U}_c$ on subcarrier $n$ and $\Lambda_{u_c c,n} = \bar{R}_{u_c c,n}$. We then concoct the following optimization problem for maximizing WSC:

$$\text{WSC}: \max_{\{X, Y, \{B_n\}\} \in \Pi} \sum_{c \in \mathcal{C}} \omega_c \sum_{n \in \mathcal{N}} y_{c,n} \sum_{u_c \in \mathcal{U}_c} x_{u_c c,n} \bar{R}_{u_c c,n}. \tag{8}$$

To specify the objective function of this optimization problem, it is necessary to derive an expression for $\bar{R}_{u_c c,n}$. In general, it is not straightforward to derive an accurate closed-form expressions for the capacity. So, we obtain the following approximations.

*Result 1:* Assume $y_{c,n} = 1$ and $x_{u_c c,n} = 1$. Then,

$$\bar{R}_{u_c c,n} \approx \int_0^\infty \frac{e^{-N_t w \sigma^2}}{w} \prod_{c' \in C_n^c / \{c\}} \frac{1}{1 + \tilde{\rho}_{uc'} P_{c',n} \delta_{uc'} w} \left( 1 - \int_0^{\delta_{u_c c,n}} \frac{f_{\sin^2(\theta_{u_c c,n})}(x)}{1 + w P_{c,n} \rho_{uc} \delta_{u_c c,n}} \, dx \right) dw. \tag{9}$$

See Appendix-A for the proof of this. The effects of quantization inaccuracies, residual interference of neighboring cells, and the cluster-size in the subcarrier are noticeable in (9). A less computationally complex approximation is given below:

*Result 2:*

$$\bar{R}_{u_c c,n} \approx \log \left( 1 + P_{c,n} \rho_{uc} \hat{\delta}_{u_c c,n} \int_0^\infty e^{-w \sigma^2 N_t} \prod_{c' \in C_n^c / \{c\}} \frac{1}{1 + \tilde{\rho}_{uc'} P_{c',n} \delta_{uc'} w} \, dw \right). \tag{10}$$

The proof can be found in Appendix-B. In (10) $\hat{\delta}_{u_c c,n}$ is a function of $\delta_{u_c c,n}$ and other system parameters including cluster-size and the number of antennas.

The LHS of Fig. 1 illustrates the capacity of a given link where only one subcarrier is considered for an illustration purpose. We also depict Result 1 and Result 2 in this figure. This figure shows that the proposed approximations are accurate. For the sake of computation, we primarily consider Result 2 in this paper as the achievable capacity, unless otherwise stated. Note that when transmission power is of high value, the capacity is not an increasing function of transmission power as the resulting signal strength is offset by the residual inter-cell interference. We will, however, defer the incorporation of power control in this system model to the future investigation.

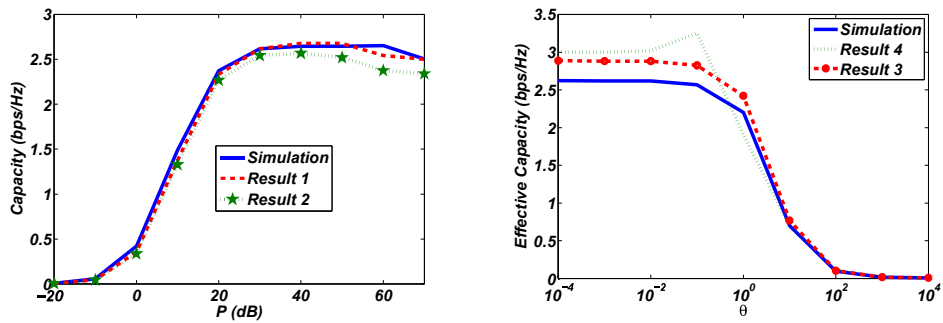

Fig. 1. LHS: Capacity of a given link $u_c c$ on subcarrier $n$ versus transmission power. RHS: The effective capacity of the link versus $\theta_{u_c}$. We consider three BSs with equal transmission power with $N_t = 5$. The allocated feedback bits are set $B_{u_c c,n} = 8$, $B_{u_c c',n} = 6$, and $B_{u_c c'',n} = 5$. We set $\alpha = 4$, and $d_{u_c c} = 300$ m, $d_{u_c c'} = 400$ m, and $d_{u_c c''} = 500$ m.

*2) Weighted Sum Effective Capacity (WSEC):* The WSC problem is usually considered in the design of CoMP systems. However, it overlooks some prominent aspects of the UE's QoS requirements, such as delay. One way to incorporate delay in the design of the RA problem is using the notion of *statistical* *delay* by *effective capacity*.[3] Introducing parameter $\theta_u$, which addresses the delay requirement of the UE $u_c$, the effective capacity, $ec_{u_c c}$, is expressed as

$$ec_{u_c c} = \frac{-1}{\theta_{u_c}} \log \left( \mathbb{E} e^{-\theta_{u_c} \sum_{n \in \mathcal{N}} x_{u_c c, n} R_{u_c c, n}} \right).$$ (11)

where $R_{u_c c, n}$ is the instantaneous transmission rate. Effective capacity is a flexible performance metric and can cover diverse performance metrics pertinent to the wireless communications systems including achievable capacity ($\theta_{u_c} \to 0$) and outage capacity ($\theta_{u_c} \to \infty$). Denoting $\bar{\mathcal{N}}_{u_c c}$ as the set of subcarriers that UE $u_c$ is active on, and recalling that subcarriers are statistically independent, one can easily confirm that

$$ec_{u_c c} = \frac{-1}{\theta_{u_c}} \log \left( \mathbb{E} e^{-\theta_{u_c} \sum_{n \in \bar{\mathcal{N}}_{u_c c}} R_{u_c c, n}} \right) = \frac{-1}{\theta_{u_c}} \log \left( \mathbb{E} \prod_{n \in \bar{\mathcal{N}}_{u_c c}} e^{-\theta_{u_c} R_{u_c c, n}} \right) = \frac{-1}{\theta_{u_c}} \log \left( \prod_{n \in \bar{\mathcal{N}}_{u_c c}} \mathbb{E} e^{-\theta_{u_c} R_{u_c c, n}} \right)$$

$$= \frac{-1}{\theta_{u_c}} \sum_{n \in \bar{\mathcal{N}}_{u_c c}} \log \left( \mathbb{E} e^{-\theta_{u_c} R_{u_c c, n}} \right) = \frac{-1}{\theta_{u_c}} \sum_{n \in \mathcal{N}} x_{u_c c, n} \log \left( \mathbb{E} e^{-\theta_{u_c} R_{u_c c, n}} \right).$$ (12)

We thus formulate the WSEC problem as:

$$\text{WSEC}: \max_{\{\boldsymbol{X}, \boldsymbol{Y}, \{\boldsymbol{B}_n\}\} \in \Pi} \sum_{c \in \mathcal{C}} \omega_c \sum_{n \in \mathcal{N}} y_{c,n} \sum_{u_c \in \mathcal{U}_c} \frac{-1}{\theta_{u_c}} x_{u_c c, n} \log \left( \mathbb{E} e^{-\theta_{u_c} R_{u_c c, n}} \right).$$ (13)

Comparing the problem WSEC with WSU in (7), we can show that the former is actually derived from the latter by setting utilization $\Lambda_{u_c c, n}$ equal to $\frac{-1}{\theta_{u_c}} \log \left( \mathbb{E} e^{-\theta_{u_c} R_{u_c c, n}} \right)$. This is an important result as the developed algorithm for WSC can also be used for WSEC by exchanging the capacity of each UE on each subcarrier with its counterpart.

Similarly to capacity, it is in general very difficult to derive an accurate closed-form expression for the effective capacity. Hence, in Appendix D we have derived the following approximation.

*Result 3:*

$$ec_{u_c c, n} \approx \frac{-1}{\theta_{u_c}} \log \left( \frac{(-1)^{N_t - |\mathcal{C}|}}{\Gamma(N_t - |\mathcal{C}|)} \int_0^\infty \frac{z^{N_t - |\mathcal{C}| - 1}}{(1 + z)^{\theta_{u_c}}} \frac{\partial^{N_t - |\mathcal{C}|} O(z)}{\partial z^{N_t - |\mathcal{C}|}} dz \right),$$ (14)

where $O(z)$ is given in Appendix D. The computational complexity of (14) is rather high. In what follows, we provide an approximation with lower computational complexity.

*Result 4:*

$$ec_{u_c c, n} \approx \frac{-1}{\theta_{u_c}} \log \left( 1 - \theta_{u_c} \bar{R}_{u_c c, n} + \frac{\theta_{u_c}^2}{2} \hat{R}_{u_c c, n} \right),$$ (15)

where $\hat{R}_{u_c c, n}$ is given by

$$\hat{R}_{u_c c_n} = \int_0^\infty \int_0^\infty \frac{e^{-\sigma^2 N_t (w_1 + w_2)}}{w_1 w_2} \prod_{c' \in \mathcal{C}_n^a / \{c\}} \frac{1}{1 + \tilde{\rho}_{u_c c'} P_{c', n} \delta_{u_c c', n}(w_1 + w_2)} (1 - \hat{K}(w_1, w_2)) dw_1 dw_2$$

---

[3]See, e.g., [47] and references therein for detailed discussion of the relationship between effective capacity and statistical delay.

and function $\hat{K}(w_1, w_2)$ is derived in Appendix E. Note that in (15), it is necessary to check if $0 < 1 - \theta_{u_c}\bar{R}_{u_cc,n} + \frac{\theta_{u_c}^2}{2}\hat{R}_{u_cc,n} \leq 1$ is valid. In our simulation, if this condition holds, we consider (15) as the effective capacity; otherwise, (14) is considered.

The RHS of Fig. 1 illustrates the effective capacity of a given link. As shown in this case, Result 3 and Result 4 also provide reasonably accurate approximations on the true effective capacity, especially when $\theta_{u_c}$ is large enough. The accuracy of the derived approximations is lower for smaller $\theta_{u_c}$. However, since for sufficiently small $\theta_{u_c}$ the effective capacity converges the capacity, one may simply consider Result 2 as an approximation of the effective capacity too. Note also that Result 3 is in general more accurate than Result 4.

Fig. 1 also reveals that by increasing $\theta_{u_c}$, the effective capacity is decreased, indicating that in limited feedback CoMP systems, the system's outage capacity can be pretty low. One way to overcome this low outage capacity is to use an effective power control mechanism tailored for cases where CDI is quantized. We leave this as our future work.

*3) Weighted Sum Energy-Efficiency (WSEE):* EE is becoming important in designing the environmentally and economically sustainable wireless systems. The methods devised for maximizing WSC may thus yield energy-inefficient solutions. So, we consider the WSEE as a main resource-allocation problem in this paper. To this end, the first step is to accurately represent power consumption for transmitting data at rate $R_{u_cc,n}$. According to the model in [11], the total power consumption includes the antenna radiated power, which is derived from the allocated power $P_{c,n}$ as $(1 + \tau_c)P_{c,n}$,[4] per-subcarrier circuitry power required for processing, mixer, synthesizer, and other units, denoted by $P_{c,n}^s$, and finally, dynamic circuitry power proportionally related to the transmitted data rate, which is $\zeta_c R_{u_cc,n}$. Thus, the expected EE of BS $c$ can be approximated as

$$(\text{e2})_c = \mathbb{E}\frac{\sum\limits_{n\in\mathcal{N}} y_{c,n}\sum\limits_{u\in\mathcal{U}_c} x_{u_cc,n} R_{u_cc,n}}{\sum\limits_{n\in\mathcal{N}} y_{c,n}\sum\limits_{u\in\mathcal{U}_c} x_{u_cc,n}(P_{c,n}^s + (1+\tau_c)P_{c,n} + \zeta_c R_{u_cc,n})}, \tag{16}$$

which is upper-bounded by

$$(\text{E2})_c = \frac{\sum\limits_{n\in\mathcal{N}} y_{c,n}\sum\limits_{u\in\mathcal{U}_c} x_{u_cc,n} \bar{R}_{u_cc,n}}{\sum\limits_{n\in\mathcal{N}} y_{c,n}\sum\limits_{u\in\mathcal{U}_c} x_{u_cc,n}(P_{c,n}^s + (1+\tau_c)P_{c,n} + \zeta_c \bar{R}_{u_cc,n})}, \tag{17}$$

due to the Jonsen's Inequality. We therefore consider $(\text{E2})_c$ as the true EE. The optimization problem that maximizes the EE is thus formulated as

$$\text{WSEE}: \max_{\{\boldsymbol{X},\boldsymbol{Y},\{\boldsymbol{B}_n\}\}\in\Pi}\sum_{c\in\mathcal{C}}\omega_c\frac{\sum\limits_{n\in\mathcal{N}} y_{c,n}\sum\limits_{u\in\mathcal{U}_c} x_{u_cc,n} \bar{R}_{u_cc,n}}{\sum\limits_{n\in\mathcal{N}} y_{c,n}\sum\limits_{u\in\mathcal{U}_c} x_{u_cc,n}(P_{c,n}^s + (1+\tau_c)P_{c,n} + \zeta_c \bar{R}_{u_cc,n})}. \tag{18}$$

---

[4] $\tau_c$ indicates the ratio of peak-to-average parameter and the drain efficiency of the radio amplifier.

Comparing problem WSEE with two previously devised optimization problems, we find that the composition of this one is different, calling for a new set of algorithms which are different from those designed for WSC and WSEC. We would like to come up with a composition that allows the usage of the same set of algorithms to control the complexity issues. To meet this goal, we consider

$$(\text{E2})_{u_c c,n} \approx \frac{\bar{R}_{u_c c,n}}{P^s_{c,n} + (1+\tau_c)P_{c,n} + \zeta_c \bar{R}_{u_c c,n}}, \tag{19}$$

as an approximate of the EE corresponding to UE $u_c$ on subcarrier $n$ of cell $c$ and reformulate WSEE as the following optimization problem:

$$\text{WSEE}: \quad \max_{\{\boldsymbol{X}, \boldsymbol{Y}, \{\boldsymbol{B}_n\}\} \in \Pi} \sum_{c \in \mathcal{C}} \omega_c \sum_{n \in \mathcal{N}} y_{c,n} \sum_{u \in \mathcal{U}_c} x_{u_c c,n} \frac{\bar{R}_{u_c c,n}}{P^s_{c,n} + (1+\tau_c)P_{c,n} + \zeta_c \bar{R}_{u_c c,n}}. \tag{20}$$

*4) Weighted Sum Effective Energy Efficiency (WSEEE):* Like capacity, EE overlooks the UE's statistical delay requirement. To compensate this, we may simply substitute capacity terms in the formulation of EE with effective capacity and define effective EE as:

$$(\text{E3})_c = \frac{\sum_{u \in \mathcal{U}_c} \frac{-1}{\bar{\theta}_{u_c}} \log \left( \mathbb{E} e^{-\theta_{u_c} \sum_{n \in \mathcal{N}} y_{c,n} x_{u_c c,n} R_{u_c c,n}} \right)}{\sum_{n \in \mathcal{N}} y_{c,n} \sum_{u \in \mathcal{U}_c} x_{u_c c,n}(P^s_{c,n} + (1+\tau_c)P_{c,n}) + \zeta_c \sum_{u \in \mathcal{U}_c} \frac{-1}{\bar{\theta}_{u_c}} \log \left( \mathbb{E} e^{-\theta_{u_c} \sum_{n \in \mathcal{N}} y_{c,n} x_{u_c c,n} R_{u_c c,n}} \right)}. \tag{21}$$

Applying (12) and following the same logic presented above, we then reformulate effective EE as

$$(\text{E3})_{u_c c,n} \approx \frac{\frac{-1}{\theta_{u_c}} \log \left( \mathbb{E} e^{-\theta_{u_c} R_{u_c c,n}} \right)}{P^s_{c,n} + (1+\tau_c)P_{c,n} + \zeta_c \frac{-1}{\theta_{u_c}} \log \left( \mathbb{E} e^{-\theta_{u_c} R_{u_c c,n}} \right)}, \tag{22}$$

and consequently propose the following optimization problem:

$$\text{WSEEE}: \quad \max_{\{\boldsymbol{X}, \boldsymbol{Y}, \{\boldsymbol{B}_n\}\} \in \Pi} \sum_{c \in \mathcal{C}} \omega_c \sum_{n \in \mathcal{N}} y_{c,n} \sum_{u \in \mathcal{U}_c} x_{u_c c,n} \frac{\frac{-1}{\theta_{u_c}} \log \left( \mathbb{E} e^{-\theta_{u_c} R_{u_c c,n}} \right)}{P^s_{c,n} + (1+\tau_c)P_{c,n} + \zeta_c \frac{-1}{\theta_{u_c}} \log \left( \mathbb{E} e^{-\theta_{u_c} R_{u_c c,n}} \right)}. \tag{23}$$

The above four proposed RA problems are instances of the general RA problem in (7): WSC ($\Lambda_{u_c c,n} = \bar{R}_{u_c c,n}$), WSEC ($\Lambda_{u_c c,n} = ec_{u_c c,n}$), WSEE ($\Lambda_{u_c c,n} = (\text{E2})_{u_c c,n}$), and WSEEE ($\Lambda_{u_c c,n} = (\text{E3})_{u_c c,n}$).

We will develop RA algorithms in the following section.

## V. Algorithms for Coordinated Resource Allocation

We now provide effective and computationally affordable solutions for problem WSR. As demonstrated in the previous section, what remains is a matter of substituting the link utilization $\Lambda_{u_c c,n}$ according to the specific RA problem. But, there are a number of challenges that make finding a solution to the problem WSU very challenging. First of all, WSU problem is a combinatorial optimization problem, and hence simple brute-force or exhaustive search won't work for the practical systems where the size of matrices $\boldsymbol{X}$, $\boldsymbol{Y}$ and $\boldsymbol{B}_n \forall n$ are large. The inherent complexity of solving (7) is exacerbated since $B_{tot}$ can, in practice, be very great that should be split among subcarriers, cells, and active UEs. In fact, the worst-case complexity of FBP can grow as fast as $(|\mathcal{N}||\mathcal{C}|^2)^{B_{tot}}$. The ultimate complexity can be much greater than this because of interplay among SA, p-sCSD and FBP. We, therefore, develop an iterative approach for solving the optimization problem WSU. Each iteration contains a number of greedy solutions to deal with FBP, p-sCSD and SA.

Let $t \geq 0$ be the iteration number; we use $\boldsymbol{X}[t]$, $\boldsymbol{Y}[t]$ and $\boldsymbol{B}_n[t] \forall n$ to represent the values at the $t$-th iteration, and the parameter $\epsilon \in (0, 1)$ to control the required accuracy for terminating the iteration. The proposed algorithm iterates as follows.

1) Set $t = 0$. Initially, assume that BS $c$ on each subcarrier $n$ is active provided that there is at least one UE associated with it. Generate matrix $\boldsymbol{X}[t]$ randomly such that only one UE stays active in each active cell and on each subcarrier.

2) Execute Algorithm gFBP to find $\{\boldsymbol{B}_n[t]\}_n$. Set $\lambda_B[t] = \lambda$.

3) Let $t \leftarrow t + 1$. Execute Algorithm C-bSA and update matrices $\boldsymbol{X}[t]$ and $\boldsymbol{Y}[t]$. Set $\lambda_X[t] = \lambda$.

4) If $|\lambda_B[t - 1] - \lambda_X[t]| \leq \epsilon$ holds, then terminate the iteration; otherwise, go to step 2.

The pseudo-code algorithms *greedy Feedback Bit Partitioning (gFBP)* and *Cluster-based SA (C-bSA)* can be found in Algorithm 1 and Algorithm 4, respectively. These algorithms will be detailed next. We start with gFBP, and then focus on C-bSA that tackles SA and p-sCSD.

*Remark 3:* $\lambda_B[0]$ is actually the WSU when the SA, UE scheduling, and p-sCSD mechanisms all are done randomly. One way to assess to what extent these mechanisms contribute to the overall network performance is by comparing $\lambda_B[0]$ and, for example, $\lambda_B[1]$. Our numerical results presented in Section VI indicate that this can be up to 7-fold.

## A. Greedy Feedback Bit Partitioning

We use the notation $\mathcal{C}_n^a[t]$ to denote the active BSs on subcarrier $n$ and iteration $t$. Further, $u_c^*[t, n]$ is the selected UE at the cell $c$ on subcarrier $n$ and iteration $t$. $B_{tot}$ available feedback bits should optimally be partitioned among the subcarriers, cells in $\mathcal{C}_n^a[t]$, and the selected UEs $u_c^*[t, n]$. Let $B_n$ be the assigned feedback bits to the subcarrier $n$ such that $\sum_{n \in \mathcal{N}} B_n = B_{tot}$. Note that Algorithm gFBP obtains $B_n$.

Constraint $C3$ dictates

$$\sum_{c \in \mathcal{C}_n^a[t]} \sum_{c' \in \mathcal{C}_n^a[t]} B_{u_c^*[t,n]c',n} = B_n, \tag{24}$$

where the double summation on BS indices is needed because 1) $B_n$ bits should be partitioned among active BSs, i.e., $c \in \mathcal{C}_n^a[t]$, and 2) in each cell $c \in \mathcal{C}_n^a[t]$ the assigned feedback bits to UE $u_c^*[t,n]$ have to be further partitioned among all interfering BSs and the home BS. Therefore, Algorithm gFBP needs to deal with the FBP in three levels including subcarriers, cells, and UEs.

Let $\dot{B}_n$ $\forall n$ be the initial assigned feedback bits to subcarriers. Instead of $\dot{B}_n = 0, \forall n$, gFBP considers the initial partitioning point as $\dot{B}_n = \lfloor \frac{B_{tot}}{\iota N} \rfloor, \forall n$, where $\iota = 1, 2, \ldots$. This provides flexibility for making a trade-off between the complexity of the algorithm and its resultant performance. Also, the complexity dramatically increases with an increase of $\iota$.

---

**Algorithm 1** Greedy Feedback Bit Partitioning (gFBP)

---

1: Input: Matrices $\boldsymbol{X}[t]$ and $\boldsymbol{Y}[t]$
2: Output: Matrices $\{\boldsymbol{B}_n[t]\}_n$ and $\lambda[t]$
3: Set $\dot{B}_n = \lfloor \frac{B_{tot}}{\iota N} \rfloor, \forall n$.               ▷ Initial feedback bits assigned to subcarriers, $\iota = 1, 2, \ldots$
4:   $\dot{B}_n = 0, \forall n$, and $\boldsymbol{b} = \{\dot{B}_1 + \ddot{B}_1, \dot{B}_2 + \ddot{B}_2 \ldots, \dot{B}_N + \ddot{B}_N\}$
5: Set $\boldsymbol{B}_n[t] = \boldsymbol{0}^{|\mathcal{U}| \times |\mathcal{C}|}, \forall n$
6: $B_{res} = B_{tot} - \sum_{n \in \mathcal{N}} \dot{B}_n$                   ▷ The available feedback bits to be shared among subcarriers
7: **while** $B_{res} > 0$ **do**
8:   **for** $n' = 1, \ldots, N$ **do**            ▷ Deciding which subcarrier is the worthiest one for adding one feedback bit
9:    Set $\boldsymbol{b}^{(n')} = \{\dot{B}_1 + \ddot{B}_1, \dot{B}_2 + \ddot{B}_2, \ldots, \dot{B}_{n'} + \ddot{B}_{n'} + 1, \ldots, \dot{B}_N + \ddot{B}_N\}$     ▷ Adding one bit to position $n'$
10:    **for** $n = 1, \ldots, N$ **do**
11:     Execute C-lFBP for $B_n = [b^{(n')}]_n$, $\{u_c^*[t,n]\} = \boldsymbol{X}[t]_{:,n}$, and $C_n^a[t] = \{c : \boldsymbol{Y}[t]_{c,n} = 1\}$   ▷ $B_n$, $\{u_c^*[t,n]\}$, and $C_n^a[t]$ are inputs of C-lFBP (see Algorithm 2)
12:     Update $\lambda_n(\boldsymbol{b}^{(n')}) = \tilde{\lambda}_n[t]$ and $\tilde{\boldsymbol{B}}_n(\boldsymbol{b}^{(n')}) = \tilde{\boldsymbol{B}}_n[t]$          ▷ $\tilde{\lambda}_n[t]$ and $\tilde{\boldsymbol{B}}_n[t]$ are outputs of C-lFBP
13:    **end for**
14:   **end for**
15:   $n'' = \max_{n'} \left( \sum_{n \in \mathcal{N}} \lambda_n(\boldsymbol{b}^{(n')}) - \sum_{n \in \mathcal{N}} \lambda_n(\boldsymbol{b}) \right)$       ▷ Subcarrier $n''$ is the worthiest one to have one extra feedback bit
16:   Set $\lambda[t] = \sum_{n \in \mathcal{N}} \lambda_n(\boldsymbol{b}^{(n'')})$, $\boldsymbol{b} = \boldsymbol{b}^{(n'')}$, $\boldsymbol{B}_n[t] = \tilde{\boldsymbol{B}}_n(\boldsymbol{b}^{(n'')})$, and $B_{res} = B_{res} - 1$     ▷ Updating parameters and outputs
17: **end while**

---

In gFBP *Cluster-level Feedback Bit Partitioning (C-lFBP)* is called. C-lFBP correspondingly partitions the assigned feedback bits to each subcarrier among active BSs in the cluster. Let $b_{c,n} = \sum_{c' \in \mathcal{C}_n^a[t]} B_{u_c^*[t,n]c',n}$ that is the assigned feedback bits to cell $c \in \mathcal{C}_n^a$ on subcarrier $n$. According to (24), it is also necessary to partition $B_n$ such that $\sum_{c \in \mathcal{C}_n^a[t]} b_{c,n} = B_n$ stays valid. Such partitioning is subject to maximization of the utilities summed over all selected UEs $u_c^*[t,n]$ in the cluster on subcarrier $n$. The details can be found in Algorithm 2.

In the body of Algorithm C-lFBP, *UE-level Feedback Bit Partitioning (U-lFBP)* is called. This algorithm is responsible for partitioning the assigned feedback bits $b_{c,n}$ to the cell $c \in \mathcal{C}_n^a[t]$ among the active BSs. The selected UE $u_c^*[t,n]$ will then use such partitioning for quantizing all the CDIs originated from

---

**Algorithm 2** Cluster-level Feedback Bit Partitioning (C-lFBP)

1:    Input: Set of selected UEs $u_c^*[t, n]$, set $\mathcal{C}_n^a[t]$, and available feedback bits to subcarrier $n$ $B_n$          $\triangleright$ $B_n$ is specified from gFBP

2:    Output: $\lambda_n[t]$ and $\boldsymbol{B}_n[t] = \boldsymbol{0}^{|\mathcal{U}| \times |\mathcal{C}|}$

3:    Set $\boldsymbol{b}_n = [\dot{b}_{c,n}, \forall c] = \boldsymbol{0}^{1 \times |\mathcal{C}|}$, $\boldsymbol{B}_n[t] = \boldsymbol{0}^{|\mathcal{U}| \times |\mathcal{C}|}$          $\triangleright$ $\boldsymbol{b}_n$ is the partitioned feedback bits $B_n$ among the active cells on subcarrier $n$

4:    Set $\boldsymbol{\lambda}_n(\boldsymbol{b}_n) = \boldsymbol{0}^{1 \times |\mathcal{C}|}$ and $\tilde{\boldsymbol{B}}_n(\boldsymbol{b}_n) = \boldsymbol{0}^{|\mathcal{U}| \times |\mathcal{C}|}$          $\triangleright$ $\boldsymbol{\lambda}_n(\boldsymbol{b}_n)$ and $\tilde{\boldsymbol{B}}_n(\boldsymbol{b}_n)$ emphasizes the dependency to vector $\boldsymbol{b}_n$

5:    **while** $B_n > 0$ **do**

6:      **for** $c' \in \mathcal{C}_n^a[t]$ **do**          $\triangleright$ Deciding which cell is the worthiest to have an extra feedback bit

7:        Set $\boldsymbol{b}_n^{(c')} = \{b_{1,n}, b_{2,n}, \ldots, b_{c',n} + 1, \ldots, b_{|\mathcal{C}|,n}\}$          $\triangleright$ Adding one extra feedback bit to cell $c' \in \mathcal{C}_n^a[t]$

8:        **for** $\forall c \in \mathcal{C}_n^a[t]$ **do**

9:          Execute U-lFBA for UE $u_c^*[t, n]$, $\mathcal{C}_n^a[t]$, and $b_{c,n} = [\boldsymbol{b}_n^{(c')}]_c$          $\triangleright$ $u_c^*[t, n]$, $\mathcal{C}_n^a[t]$, and $b_{c,n}$ are the input of U-lFBP (see Algorithm 3)

10:         Set $[\boldsymbol{\lambda}_n(\boldsymbol{b}_n^{(c')})]_c = \hat{\Lambda}_{u_c^*[t,n]c,n}$ and $[\tilde{\boldsymbol{B}}_n^{(c')}(\boldsymbol{b}_n^{(c')})]_{u_c^*[t,n],:} = \tilde{\boldsymbol{b}}_{u_c^*[t,n]c,n}$          $\triangleright$ $\hat{\Lambda}_{u_c^*[t,n]c,n}$ and $\tilde{\boldsymbol{b}}_{u_c^*[t,n]c,n}$ are output of U-lFBP

11:        **end for**

12:      **end for**

13:      $\hat{c} = \max_{c'} \left( \sum_{c \in \mathcal{C}_n^a[t]} [\boldsymbol{\lambda}_n(\boldsymbol{b}_n^{(c')})]_c - \sum_{c \in \mathcal{C}_n^a[t]} [\boldsymbol{\lambda}_n(\boldsymbol{b}_n)]_c \right)$          $\triangleright$ Cell $\hat{c}$ is the worthiest cell to have one extra feedback bit on subcarrier $n$

14:      Update $\lambda_n[t] = \sum_{c \in \mathcal{C}_n^a[t]} [\boldsymbol{\lambda}_n(\boldsymbol{b}_n^{(\hat{c})})]_c$, $B_n = B_n - 1$, $\boldsymbol{B}_n[t] = \tilde{\boldsymbol{B}}_n^{(\hat{c})}(\boldsymbol{b}_n^{(\hat{c})})$, and $\boldsymbol{b}_n = \boldsymbol{b}_n^{(\hat{c})}$          $\triangleright$ Updating outputs and parameters

15:    **end while**

---

BSs $c \in \mathcal{C}_n^a[t]$ (including home cell $c$ and interfering BSs). A higher share of available feedback bits is considered for a BS that ultimately contributes more to the improvement of $\Lambda_{u_c^*[t,n]c,n}[t]$. U-lFBP is detailed in Algorithm 3.

---

**Algorithm 3** *UE-level Feedback Bit Partitioning* (U-lFBP)

1:    Input: UE $u_c^*[t, n] \in \mathcal{U}_c$ and set $\mathcal{C}_n^a[t]$ on subcarrier $n$ and iteration $t$, $b_{c,n}$ bits available for partitioning

2:    Output: $\Lambda_{u_c^*[t,n]c,n}[t]$ and $\boldsymbol{b}_{u_c^*[t,n]c,n}^{1 \times |\mathcal{C}|}[t]$          $\triangleright$ $\boldsymbol{b}_{u_c^*[t,n]c,n}$ is the vector indicates the partitioned feedback bits $b_{c,n}$ among the BSs at UE $u_c^*[t, n]$

3:    Set $\boldsymbol{B}_{u_c^*[t,n]c,n} = 0$, $\forall c$, $\boldsymbol{b}_{u_c^*[t,n]c,n} = \{B_{u_c^*[t,n]1,n}, \ldots, B_{u_c^*[t,n]|\mathcal{C}|,n}\}$

4:    **while** $b_{c,n} > 0$ **do**

5:      **for** $c' \in \mathcal{C}_n^a[t]$ **do**          $\triangleright$ Deciding which BS is worthy to have an extra feedback bit for the quantization of its CDI

6:        Set $\boldsymbol{b}^{(c')} = \{B_{u_c^*[t,n]1,n}, \ldots, B_{u_c^*[t,n]c',n} + 1, \ldots, B_{u_c^*[t,n]|\mathcal{C}|,n}\}$          $\triangleright$ Assigning one extra feedback bit to the BS $c' \in \mathcal{C}_n^a[t]$

7:        Calculate $\Lambda_{u_c^*[t,n]c,n}$ for $\boldsymbol{b}^{(c')}$ and update $[\Lambda_{u_c^*[t,n]c,n}(\boldsymbol{b}^{(c')})]_{c'} = \Lambda_{u_c^*[t,n]c,n}$

8:      **end for**

9:      Find $\hat{c} = \max_{c'} \left( [\Lambda_{u_c^*[t,n]c,n}(\boldsymbol{b}^{(c')})]_{c'} - [\Lambda_{u_c^*[t,n]c,n}(\boldsymbol{b})]_{c'} \right)$          $\triangleright$ One extra feedback bit is assigned to BS $c'$

10:      Update $\Lambda_{u_c^*[t,n]c,n}[t] = [\Lambda_{u_c^*[t,n]c,n}(\boldsymbol{b}^{(\hat{c})})]_{\hat{c}}$, $\boldsymbol{b}_{u_c^*[t,n]c,n}[t] = \boldsymbol{b}^{(\hat{c})}$, $b_{c,n} = b_{c,n} - 1$          $\triangleright$ Updating outputs and parameters

11:    **end while**

---

*Remark 4:* Similar to Algorithm U-lFBP, a greedy algorithm was previously developed in [43] for limited-feedback interference channels. Nevertheless, in [43], the available feedback bits for each UE were given, i.e., obtaining $b_{c,n}$ $\forall c, n$ is not the job of FBP in our system model. Furthermore, the system setup there is for a single carrier. In contrast, we extended the greedy FBP into levels of subcarriers and cells beside UEs, so obtaining $b_{c,n}$ $\forall c, n$ is also the job of FBP.

## B. SA, UE Scheduling, and Cluster-Size Determination: Algorithm C-bSA

After specifying matrices $\boldsymbol{B}_n[t-1]$ $\forall n$, we may have to re-evaluate the matrices $\boldsymbol{X}[t]$ and $\boldsymbol{Y}[t]$. This is taken care of via C-bSA, see Algorithm 4. C-bSA determines which BSs need to stay active on each subcarrier. There is a counter, denoted by $l = 0, 1, \ldots, |\mathcal{C}| - 1$, that determines the number of BSs required to be switched off on each specific subcarrier. The algorithm further selects those $l$ BSs that should be switched off. Moreover, on each subcarrier $n$, C-bSA determines which UE has to be selected in each active cell. Note that from gFBP, we have an estimate of $B_n$. However, the previously founded matrices $B_n[t-1]\ \forall n$ may not be valid any longer as the algorithm determines whether there are better UEs to be selected on each subcarrier. It is then necessary to redistribute $B_n$ according to SA and p-sCSD by calling algorithms C-lFBP and U-lFBP.

---

**Algorithm 4** C-bSA Algorithm

---

1:    Input: $\boldsymbol{B}_n[t-1]\ \forall n$, $\boldsymbol{X}[t-1]$, and $\boldsymbol{Y}[t-1]$

2:    Output: Matrices $\boldsymbol{X}[t]$ and $\boldsymbol{Y}[t]$

3:    Set $\boldsymbol{X}[t] = \mathbf{0}$ and $\boldsymbol{Y}[t] = \mathbf{0}$

4:    **for** $n = 1, 2, \ldots, N$ **do**

5:      Set $l = 0$, $\mathcal{C}_{l,n} = \{c : [\boldsymbol{Y}[t-1]]_{c,n} = 1\}$           ▷ $l$ counts the BSs that should be switched off. Respectively, set $\mathcal{C}_{l,n}$ contains $|\mathcal{C}| - l$ active BS on subcarrier $n$

6:      **for** $c \in \mathcal{C}_{l,n}$ **do**           ▷ Assuming partitioning $\boldsymbol{B}_n[t-1]$ is valid, check to see if it is possible to select worthier UEs in the cluster

7:          $b_{c,n} = \sum_{u \in \mathcal{U}_c} \boldsymbol{B}_n(:,c)[t-1]$

8:          **for** $u \in \mathcal{U}_c$ **do**

9:              Execute U-lFBA for $u_c^*[t,n] = u$ and $b_{c,n}$, and accordingly update $\Lambda_{uc,n}$

10:          **end for**

11:          Find $u_c^*[t,n] = \arg\max_{u \in \mathcal{U}_c} \Lambda_{uc,n}^*$ and set $[\boldsymbol{X}[t]]_{u_c^*[t,n],n} = 1$       ▷ Deciding UEs that should be active in each cell on subcarrier $n$ and updating $\boldsymbol{X}[t]$ respectively

12:      **end for**

13:      Calculate $\lambda_{l,n} = \sum_{c \in \mathcal{C}_{l,n}} \Lambda_{u_c^*[t,n]c,n}$          ▷ $\lambda_{l,n}$ is the cluster utilization on subcarrier $n$ when $l$ BSs are switched off

14:      Set $\hat{\lambda}_{l,n} = 0$, $\boldsymbol{B}_{l,n} = \mathbf{0}$, $\dot{\boldsymbol{X}}[t] = \mathbf{0}$, and $l = l+1$       ▷ $\boldsymbol{B}_{l,n}$ tracks the feedback bit partitioning in the cluster on subcarrier $n$ when $l$ BSs are switched off

15:      **while** $l < |\mathcal{C}|$ **do**          ▷ Checking how many BSs should be switched off on subcarrier $n$

16:          Find $\hat{c}_{l,n} = \arg\min_{c \in \mathcal{C}_{l,n}} \Lambda_{u_c^*[t,n]c,n}$          ▷ Finding BS $\hat{c}_{l,n}$ that offers smallest utilization among other active BSs

17:          Let $\hat{\mathcal{C}}_{l,n} = \mathcal{C}_{l,n}/\{\hat{c}_{l,n}\}$          ▷ Temporarily switching off BS $\hat{c}_{l,n}$

18:          Execute C-lFBP for $C_n^a[t] = \hat{\mathcal{C}}_{l,n}$ and $B_n = \sum_{c \in \mathcal{C}} b_{c,n}$. Accordingly, update $\boldsymbol{B}_{l,n}$       ▷ Since $C_n^a[t]$ is updated it is necessary to update $\boldsymbol{B}_{l,n}$

19:          **for** $c \in \hat{\mathcal{C}}_{l,n}$ **do**          ▷ Checking to confirm whether or not switching off BS $\hat{c}_{l,n}$ is beneficial

20:              $b_{l,c,n} = \sum_{u \in \mathcal{U}_c} \boldsymbol{B}_{l,n}(:,c)$

21:              **for** $u \in \mathcal{U}_c$ **do**          ▷ Finding the worthiest UE in each active cell based on updated $\boldsymbol{B}_{l,n}$ and $C_n^a[t]$

22:                  Execute U-lFBA for $u_c^*[t,n] = u$ and $b_{l,c,n}$, and accordingly update $\Lambda_{uc,n}$ and $\hat{b}_{uc,n}$

23:              **end for**

24:          Find $u_c^*[t,n] = \arg\max_{u \in \mathcal{U}_c} \Lambda_{uc,n}$ and update $[\boldsymbol{B}_{l,n}]_{(u_c^*[t,n],:)} = \hat{b}_{u_c^*[t,n]c,n}$. Set $[\dot{\boldsymbol{X}}[t]]_{u_c^*[t,n],n} = 1$

25:          **end for**

26:          Calculate $\hat{\lambda}_{l,n} = \sum_{c \in \hat{\mathcal{C}}_{l,n}} \Lambda_{u_c^*[t,n]c,n}$.

27:          **if** $\lambda_{l,n} < \hat{\lambda}_{l,n}$ **then**

28:              Set $\mathcal{C}_{l,n} = \hat{\mathcal{C}}_{l,n}$, $\lambda_{l,n} = \hat{\lambda}_{l,n}$, $[\boldsymbol{X}[t]]_{(:,n)} = [\dot{\boldsymbol{X}}[t]]_{(:,n)}$, $[\boldsymbol{Y}[t]]_{(\hat{c}_{l,n},n)} = 0$, and $l = l+1$       ▷ Switching off BS $\hat{c}_{l,n}$ is valid. Updating parameters accordingly.

29:          **else**

30:              $l = |\mathcal{C}|$.          ▷ Switching off BS $\hat{c}_{l,n}$ is detrimental. No need to go further.

31:          **end if**

32:      **end while**

33:      Set $C_n^a[t] = \mathcal{C}_{l,n}$, $\boldsymbol{B}_n[t] = \boldsymbol{B}_{l,n}$ and $\lambda_n = \lambda_{l,n}$.

34:    **end for**

35:    Update $\lambda[t] = \sum_{n \in \mathcal{N}} \lambda_n$, $\boldsymbol{Y}[t]$, and $\boldsymbol{X}[t]$.

---

*Remark 5:* To implement the proposed RA algorithms in this paper the central processor merely needs to know path-loss attenuations of each UE with respect to all BSs in the cluster. After specifying SA, UE scheduling, and FBP, BSs will respectively inform the selected UEs on each subcarrier with available feedback resources for CDI quantization/feedback during the entire communication session.

*Remark 6:* Comparing to the cases in which U-lFBP and/or C-lFBP are pre-specified—for instance, by equally partitioning, respectively, $B_n$ among the active cells and $B_n/|\mathcal{C}_n^a|$ among the interfering and attending links—the proposed gFBP algorithm imposes higher computational complexity and signaling overhead. Note that central processors are highly powerful and can easily handle the imposed computational complexity. Moreover, the imposed overhead may not be a crushing issue as we stated in Remark 5, the developed RA policy in this paper, and consequently devised FBP, stays valid provided that UEs do not experience a substantial path-loss changes. This can be granted particularly when UEs are rather slowly moving with respect to the time scale of RA execution. For the cases that some UEs experience considerable path-loss attenuations, perhaps due to shadowing and/or high mobility, it in then up to the system designer to trade-off signaling overhead and the performance.

## VI. SIMULATION RESULTS

Here we study the performance of the developed algorithms for WSC, WSEC, WSEE, and finally, WSEEE. The simulation setup is depicted in Fig. 2. The cluster under consideration consists of $|\mathcal{C}|$ BSs. In each cluster, BSs are positioned in a circle centered at the cluster-center, where the central unit processor is located, with radius $0 \leq D \leq R$ where $R$ is the cluster radius. We set $R$ to 1000 meters. The rest of the parameters are $|\mathcal{N}| = 64$, $P_c = 10$ W $\forall c$, $\sigma^2 = 10^{-10}$ W, $\alpha = 4$. Processing power is $0.5$ W that is equally divided among all of the active subcarriers. Moreover, $\tau_c = \zeta_c = 0.1$ and $\omega_c = 1/|\mathcal{C}|$ $\forall c$. Finally, for all UEs we set $\theta_u = 1$. We randomly position UEs in the cluster. Let UEs simply be associated with the closest BS.

To model inter-cluster interference, we consider the first and second tiers of the clusters surrounding the considered cluster. The effect of inter-cluster interference is then captured through the AWGN. To do so, the fading fluctuation is substituted with its average value that is $N_t$. As a result, the variance of AWGN mainly depends on the position of the UEs in the considered cluster with respect to the all active BSs in the interfering clusters. The power level at the BSs in the interfering clusters is specified by their on/off status by applying the proposed algorithms in this paper, adopting wrap-around technique.

*a)* **How important is it to optimally assign subcarriers and determine the cluster-size?:** To assess the advantages of SA coupled with p-sCSD, first we need to investigate the convergence of the proposed algorithm in this paper. This is because, as discussed in Remark 2, in the first iteration $t = 0$, all the BSs are turned on on all subcarriers and UEs are scheduled randomly. Hence, comparing the performance of the network at iteration 0 and the next iteration, after executing C-bSA algorithm, reveals the benefits of SA and p-sCSD.

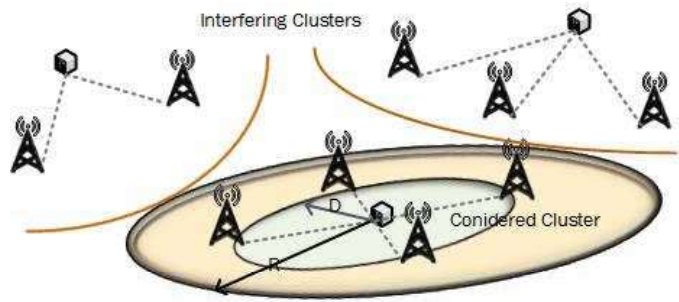

Fig. 2.   The network setting used for simulation. BSs in each cluster are connected to the central processor and are $D$ meters away from the processor. The radius of each cluster is $R$ meters.

Fig. 3 demonstrates the convergence property of the developed algorithm in this paper. As it is seen, in all cases, almost after 2 iterations the algorithms converge. Here we set $\epsilon = 0.1$. More importantly, as also mentioned above, these illustrations indicate the effects of SA coupled with p-sCSD on the improvement of the network performance. For example, for the case of WSC and when $N_t = 8$, $|\mathcal{U}| = 50$, it is seen that C-bCA algorithm results in a promising 4.5-fold performance improvement. This performance improvement is further escalated to 7-fold when $|\mathcal{U}| = 200$. This confirms the C-bCA algorithm in effectively exploiting the multi-user, multi-carrier and multi-cell diversities in McLf-CoMP.

Fig. 3 also shows that for fixed number of UEs, and given $B_{tot}$, increasing number of transmit antennas from $N_t = 8$ to $N_t = 12$ and $N_t = 16$, results in a slight degradation in the gain corresponding to the UE scheduling and p-sCSD. This is mainly due to the fact that for larger antenna arrays, CDI inaccuracies have higher negative impact through both reducing the attending signal strengths, and weakening the capability of BSs to effectively pre-cancel the intra-cluster interference. As a result, increasing $N_t$ has to be considered jointly with available feedback capacity in the uplink channel.

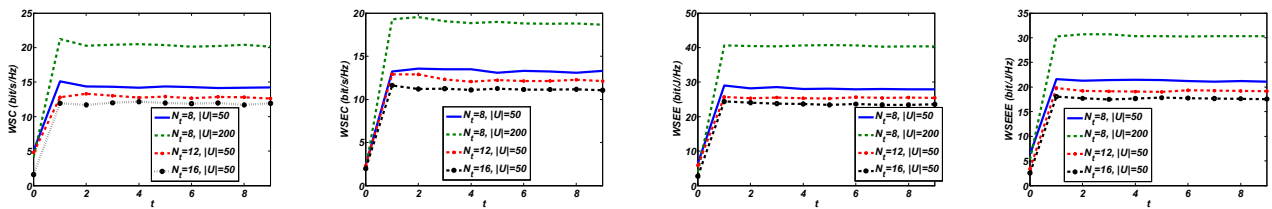

Fig. 3.   Convergence of proposed RA algorithm for all cases of WSC, WSEC, WSEE, and WSEEE when $|\mathcal{C}| = 4$.

*b)* **How to distribute BSs in a cluster?:** Fig. 4 shows the impact of distance $D$ and the cluster size $|\mathcal{C}|$ on the performance of McLf-CoMP systems. Fig. 4 suggests that there is an optimal distance $D$ at which BSs should be positioned to yield the maximum performance in all cases. In our model, this optimal distance is measured to be almost 300m. Interestingly, even for this simple setup, appropriate positioning of BSs increases the performance of the systems by up to 5x. Combined with C-bSA, Fig. 3 shows that the network performance can improve by at least an order-of-magnitude.

Fig. 4 also provides important practical insights on the cluster size, $|\mathcal{C}|$. As it is seen, increasing the number of BSs is beneficial in all cases. For example, in the case of WSC and when $D = 300$m, increasing the cluster size from 3 to 5 improves the associated performance by more than 3x. More details on the best cluster size will be given at the end of this section.

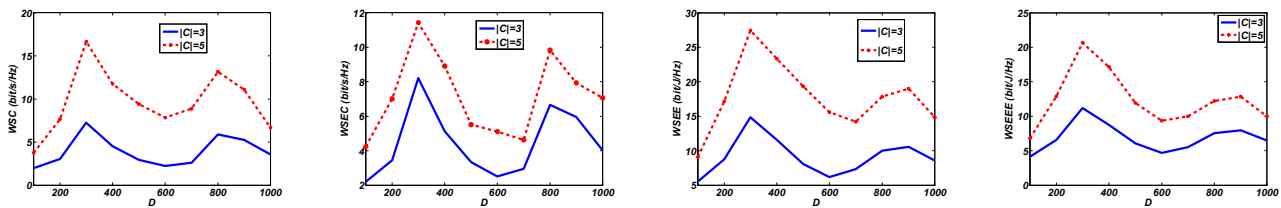

Fig. 4. Performance of proposed RA algorithm for all cases of WSC, WSEC, WSEE, and WSEEE *vs.* $D$.

*c)* **What are the benefits of partitioning feedback bits optimally?:** Here we compare the network utility based on the RA algorithm presented in this paper, referred to as *Opt.*, with that of two following systems. The first system is referred to as *Min.*, in which only Algorithm U-lFBP is modified based on minimizing the residual intra-cell interference. The second system, referred to as *Equ.*, in which the feedback resources are equally shared at each UE.

As is is seen in Fig. 5, by increasing $B_{tot}$, as expected, the network utilization is higher. We also study the impact of $\iota$ on the performance of all three systems. Note that parameter $\iota$ in fact controls the initial feedback bits assigned to the subcarriers in Algorithm gFBP. Therefore, $\iota = 1$ means that all subcarriers are initially receiving the same share of $B_{tot}$. On the other hand, increasing $\iota$ results in sharing a fraction of $B_{tot}$, which is $B_{tot}/\iota$, equally among all subcarriers. Therefore, the higher the $\iota$, the higher is the computational complexity. Increasing $\iota$ is shown to yield higher performance in all cases of different systems and different utilizations. Thus, there is a tradeoff between the complexity and performance. Nevertheless this trade off is more rewarding in the case of system *Opt.*, where increasing $\iota$ has higher impact on increasing the network utilization for larger values of $B_{tot}$.

Finally, Fig. 5 confirms that the proposed algorithm achieves a much higher performance comparing to the other two systems. It is important to note that since FBP is also incorporated in the body of Algorithm C-bSA, besides Algorithm gFBP, the advantages of proposed FBP are very promising: more than $30\%$ higher than *Equ.* system, and $20\%$ more than the *Min.* system.

We like also pointing out that our simulation results, though they are not presented here dute to space limitation, show that comparing the performance of system *Opt.* with that of the case in which Algorithm C-lFBP is modified so that to equally share the assigned feedback bits $B_n$ among the active cells (while U-lFBP is kept the same as Algorithm 3) has up to $30\%$ performance improvement. This, on the other hand, indicates the great advantages of coordinately partitioning of feedback resource among the cells belong to the cluster.

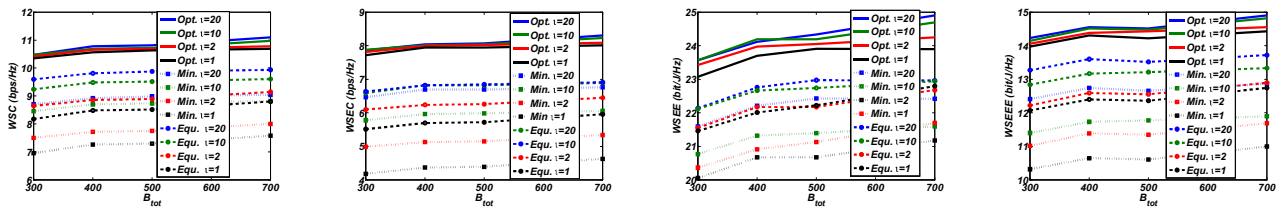

Fig. 5. Performance of the proposed RA algorithm for all cases of WSC, WSEC, WSEE, and WSEEE *vs.* $B_{tot}$ for different values of $\iota$. In this figure, we compare the results of this paper (referred to as *Opt.*) with two other systems in which U-lFBP is done respectively based on minimizing the residual intra-cell interference (referred to as *Min.*) as well as equally sharing the feedback resources (referred to as *Equ.*). Here we set $|\mathcal{C}| = 3$.

*d)* **Impact of system parameters:** We study the impact of various system parameters, including transmission power, number of antennas, number of subcarriers, and cluster size on the network utilization in Fig. 6. First, let us focus on the impact of transmission power. As the leftmost part of Fig. 6 ($|\mathcal{C}| = 3$) indicates increasing the transmission power of BSs has distinct impact on network utilization. For the case of WSC and/or WSEC, we find that the network utilization grows almost linearly with the log-scale transmission power. Note that for higher power saturation phenomenon is prevalent for CoMP systems [19], which is not presented here due to space limitation. On the other hand, as it is seen for the case of WSEE/WSEEE, there is an optimal transmission level that yields optimal network utilization. This figure further confirms that strategies yielding optimal WSEE and/or WSEEE are not necessarily optimal from the perspective of WSC and/or WSEC.

The second leftmost part of Fig. 6 ($|\mathcal{C}| = 2$) shows the impact of the number of antennas on the network utilization. In general, increasing the number of antennas results in performance reduction, albeit in some cases very slightly. Consequently, increasing the number of antennas without increasing the feedback capacity may not be a good choice. This is an important conclusion in particular for the cases where feedback traffic consumes the uplink bandwidth thus degrades the uplink performance.

The second rightmost part of Fig. 6 ($|\mathcal{C}| = 3$) shows the impact of the number of subcarriers on WSU. What we have shown so far is that by increasing $|\mathcal{N}|$ WSU is respectively increased where the pace of increment is reduced for large enough $|\mathcal{N}|$, i.e., $|\mathcal{N}| > 32$. In fact, when $|\mathcal{N}| = 32$ the system already is capable of effectively exploiting the frequency and multi-user diversities.

Finally, the rightmost part of this figure exhibits insightful results on the best cluster size. Intuitively, adding extra BSs in the cluster results in higher network utilization. However, the performance jump due to increase of the number of BSs from 3 to 5 is much greater than the case in which $|\mathcal{C}|$ is increased from 5 to 7. This is an importance design guideline to know that $|\mathcal{C}| = 5$ is almost as good as $|\mathcal{C}| = 7$, since it allows the designer to save capital investment associated with installing extra BSs and the associated backhauls as well as electrical bills without degrading the network utilization.

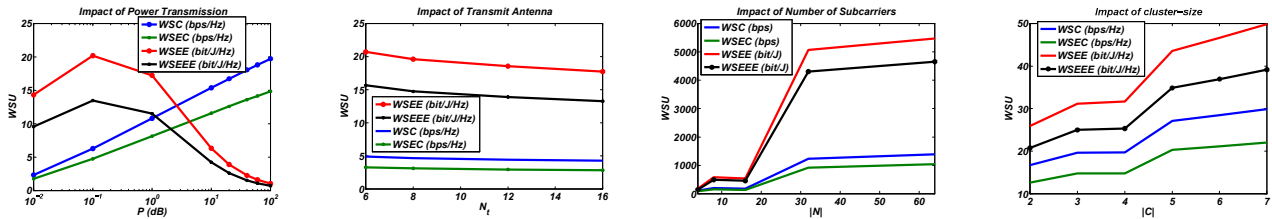

Fig. 6.   Impact of different system parameters on WSU.

## VII. Conclusions

In this paper we have studied several RA problems pertinent to the multi-carrier limited-feedback coordinated multi-point (McLf-CoMP) systems. Each cluster consisted of a number of BSs connected to the central processor and each other over the backhauls. We first constructed a general RA problem called *weighted sum utility* (WSU), and then demonstrated how it capable of being recast into optimization problems aiming at maximizing weighted sum capacity (WSC), weighted sum effective capacity (WSEC), weighted sum energy efficiency (WSEE), and weighted sum effective energy-efficiency (WSEEE). The solution of each RA problem indicated the optimal UE selection, subcarrier assignment (SA), per-subcarrier cluster-size determination (ps-CSD), and finally, feedback bit partitioning (FBP) across subcarriers, cells, and UEs. We derived analytical bounds on the capacity and the effective capacity of the coordinated system incorporating the characteristics of CDI quantization. Other performance metrics including WSEE as well as WSEEE were determined. Since all the considered RA problems were recast in a unified composition of WSU, the applicability of the same set of algorithms for solving all four constructed RA problems was possible. Nevertheless, the proposed RA problems were very complex due mainly to their combinatoric characteristics, sheer size of the optimization problems, and interactions among the optimization variables. To tackle the complexity, various greedy solutions were proposed. Specifically, a three-level greedy FBP procedure referred to as gFBP was devised for partitioning the feedback bits among the subcarriers, active cells, and active UEs. Furthermore, Algorithm C-bSA was suggested to deal with UE scheduling, SA, and p-sCSD while taking into account the characteristics of CDI quantization. The developed iterative solution was fast and shed light on several important system parameters and network optimization.

<div align="center">REFERENCES</div>

## Appendix

### A. Proof of Result 1

Let $\mathcal{C}_n^a$ be the active BSs on subcarrier $n$. We assume $c \in \mathcal{C}_n^a$. Regarding the SINR expression (2), the maximum achievable data rate of UE $u_c \in \mathcal{U}_c$ on subcarrier $n$, $\bar{R}_{u_c c, n}$, is equal to

$$\bar{R}_{u_c c,n} \;=\; \mathbb{E}\log\left(1 + \gamma_{u_c c,n}\right) = \mathbb{E}\log\left(1 + \frac{\rho_{u_c c} P_{c,n}\|\boldsymbol{h}_{u_c c,n}\|^2|\tilde{\boldsymbol{h}}_{u_c c,n}^{\dagger}\hat{\boldsymbol{f}}_{c,n}|^2}{\sigma^2 N_t + \sum\limits_{c' \in \mathcal{C}_n^a/\{c\}} \tilde{\rho}_{u_c c'} P_{c',n}\|\boldsymbol{g}_{u_c c',n}\|^2|\hat{\boldsymbol{g}}_{u_c c',n}^{\dagger}\hat{\boldsymbol{f}}_{c',n}|^2}\right). \tag{25}$$

Let random variable $J_{u_c c',n}$ denote $J_{u_c c',n} = \|\boldsymbol{g}_{u_c c',n}\|^2|\hat{\boldsymbol{g}}_{u_c c',n}^{\dagger}\hat{\boldsymbol{f}}_{c',n}|^2$ and let random variable $K_{u_c c,n}$ denote $K_{u_c c,n} = \|\boldsymbol{h}_{u_c c,n}\|^2|\tilde{\boldsymbol{h}}_{u_c c,n}^{\dagger}\hat{\boldsymbol{f}}_{c,n}|^2$.. Decomposing vector $\tilde{\boldsymbol{g}}_{u_c c'}$ as $\tilde{\boldsymbol{g}}_{u_c c'} = \cos(\theta_{u_c c'})\hat{\boldsymbol{g}}_{u_c c'} + \sin(\theta_{u_c c'})\boldsymbol{v}_{u_c c'}$, where $\boldsymbol{v}_{u_c c'}$ is perpendicular to $\hat{\boldsymbol{g}}_{u_c c'}$ and $\theta_{u_c c'}$ is the angle between $\tilde{\boldsymbol{g}}_{u_c c'}$ and $\hat{\boldsymbol{g}}_{u_c c'}$, and adopting the developed theory of QCA quantization from [37, 38], $J_{u_c c',n}$ is distributed as $J_{u_c' c',n} \propto \|\boldsymbol{g}_{u_c' c',n}\|^2 \sin^2(\theta_{u_c c',n})\mathrm{B}(1, N_t - 2)$ in which $\mathrm{B}(1, N_t - 2)$ indicates a beta random variable with parameters $1$ and $N_t - 2$. Utilizing the results of [51, 52] $J_{u_c c',n}$ is an exponentially distributed random variable where parameter $\delta_{u_c c',n}$. Note that random variables $K_{u_c c,n}$ and $J_{u_c c',n} \;\forall c' \neq c$ are independent since the direct and interfering channel vectors are independent as well as the respective quantization code-books are constructed separately. According to the rate-splitting equality [53], we rewrite (25) as:

$$\bar{R}_{u_c c,n} = \mathbb{E}\log\left(1 + \frac{\rho_{u_c c} P_{c,n} K_{u_c c,n} + \sum\limits_{c' \in \mathcal{C}_n^a/\{c\}} \tilde{\rho}_{u_c c'} P_{c',n} J_{u_c c',n}}{\sigma^2 N_t}\right) - \mathbb{E}\log\left(1 + \frac{\sum\limits_{c' \in \mathcal{C}_n^a/\{c\}} \tilde{\rho}_{u_c c'} P_{c',n} J_{u_c c',n}}{\sigma^2 N_t}\right). \tag{26}$$

Substituting the following identity $\log(1 + x) = \int_0^\infty \frac{e^{-w}}{w}(1 - e^{-wx})dw$, into (26), and then applying some manipulations, we get

$$\bar{R}_{u_c c,n} = \int_0^\infty \frac{e^{-N_t w \sigma^2}}{w} \mathbb{E}e^{-w \sum\limits_{c' \in \mathcal{C}_n^a/\{c\}} \tilde{\rho}_{u_c c'} P_{c',n} J_{u_c c',n}} \left(1 - \mathbb{E}e^{-w P_{c,n} \rho_{u_c c} K_{u_c c,n}}\right) dw. \tag{27}$$

For fixed $w$, it is straightforward to show that

$$\mathbb{E}e^{-w \sum\limits_{c' \in \mathcal{C}_n^a/\{c\}} P_{c',n} \tilde{\rho}_{u_c c'} J_{u_c c',n}} = \prod_{c' \in \mathcal{C}_n^a/\{c\}} \mathbb{E}e^{-w P_{c',n} \tilde{\rho}_{u_c c'} J_{u_c c',n}} = \prod_{c' \in \mathcal{C}_n^a/\{c\}} \frac{1}{1 + \tilde{\rho}_{u_c c'} P_{c',n} \delta_{u_c c',n} w}, \tag{28}$$

where in the last step we used the fact that $J_{u_c c',n}$ is an exponential random variable with parameter $\delta_{u_c c'}$. Substituting (28) into (27), we have

$$\bar{R}_{u_c c,n} = \int_0^\infty \frac{e^{-N_t w \sigma^2}}{w} \prod_{c' \in \mathcal{C}_n^a/\{c\}} \frac{1}{1 + \tilde{\rho}_{u_c c'} P_{c',n} \delta_{u_c c',n} w} \left(1 - \mathbb{E}e^{-w P_{c,n} \rho_{u_c c} K_{u_c c,n}}\right) dw. \tag{29}$$

Now, we calculate $\mathbb{E}e^{-w P_{c,n} \rho_{u_c c} K_{u_c c,n}}$ for fixed $w$. We expand $\tilde{\boldsymbol{h}}_{u_c c,n}$ as $\tilde{\boldsymbol{h}}_{u_c c,n} = \cos(\theta_{u_c c,n})\hat{\boldsymbol{h}}_{u_c c,n} + \sin(\theta_{u_c c,n})\boldsymbol{s}_{u_c c,n}$, where $\boldsymbol{s}_{u_c c,n}$ is orthogonal to $\hat{\boldsymbol{h}}_{u_c c,n}$ and $\theta_{u_c c,n}$ is the angle between $\tilde{\boldsymbol{h}}_{u_c c,n}$ and $\hat{\boldsymbol{h}}_{u_c c,n}$. We then apply the following approximation:

$$K_{u_c c,n} \approx \cos^2(\theta_{u_c c,n})\|\boldsymbol{h}_{u_c c,n}\|^2|\hat{\boldsymbol{h}}_{u_c c,n}^{\dagger}\hat{\boldsymbol{f}}_{c,n}|^2 + \sin^2(\theta_{u_c c,n})\|\boldsymbol{h}_{u_c c,n}\|^2|\boldsymbol{s}_{u_c c,n}^{\dagger}\hat{\boldsymbol{f}}_{c,n}|^2. \tag{30}$$

Random variable $|\boldsymbol{s}_{u_cc,n}^\dagger \hat{\boldsymbol{f}}_{c,n}|^2$ is beta with parameters $1$ and $N_t - 2$ [37], thus $\mathbb{E}|\boldsymbol{s}_{u_cc,n}^\dagger \hat{\boldsymbol{f}}_{c,n}|^2 = 1/(N_t - 1)$. Besides, random variable $\sin^2(\theta_{u_cc,n})\|\boldsymbol{h}_{u_cc,n}\|^2 |\boldsymbol{s}_{u_cc,n}^\dagger \hat{\boldsymbol{f}}_{c,n}|^2$ is exponentially distributed with parameter $\delta_{u_cc,n}$. On the other hand, as it is shown in [34] $\|\boldsymbol{h}_{u_cc,n}\|^2 |\hat{\boldsymbol{h}}_{u_cc,n}^\dagger \hat{\boldsymbol{f}}_{c,n}|^2 \sim \chi^2_{2(N_t - |\mathcal{C}_n^a| + 1)}$, which is a Chi-squared random variable with $2N_t - 2|\mathcal{C}_n^a|$ degrees-of-freedom. Our derivations suggest the following approximation

$$\mathbb{E}e^{-wP_{c,n}\rho_{u_cc}\|\boldsymbol{h}_{u_cc,n}\|^2 |\tilde{\boldsymbol{h}}_{u_cc,n}^\dagger \hat{\boldsymbol{f}}_{c,n}|^2} \approx \mathbb{E}e^{-wP_{c,n}\rho_{u_cc}\cos^2(\theta_{u_cc,n})\chi^2_{2(N_t - |\mathcal{C}_n^a|)}} \mathbb{E}e^{-wP_{c,n}\rho_{u_cc}\delta_{u_cc,n}\chi^2_2}, \tag{31}$$

in which for mathematical tractability we have assumed that random variable $\cos^2(\theta_{u_cc,n})\chi^2_{2(N_t - |\mathcal{C}|)}$ and $\delta_{u_cc,n}\chi^2_2$ are independent, which clearly are not. Combining (31) and (29), the following approximation on the achievable capacity can be suggested

$$\bar{R}_{u_cc,n} \approx \int_0^\infty \frac{e^{-N_t w \sigma^2}}{w} \prod_{c' \in \mathcal{C}_n^a / \{c\}} \frac{1}{1 + \tilde{\rho}_{u_cc'} P_{c',n} \delta_{u_cc',n} w} \left(1 - \frac{\mathbb{E}e^{-wP_{c,n}\rho_{u_cc}\cos^2(\theta_{u_cc,n})\chi^2_{2(N_t - |\mathcal{C}_n^a|)}}}{1 + \rho_{u_cc} P_{c,n} \delta_{u_cc,n} w}\right) dw. \tag{32}$$

By substituting QCA pdf (3) in (32), the desired result is then obtained.

## B. Proof of Result 2

We apply Jensen's inequality to obtain $\bar{R}_{u_cc,n} \leq \log\left(1 + \mathbb{E}\gamma_{u_cc,n}\right)$. It is then enough to provide an expression for $\mathbb{E}\gamma_{u_cc,n}$. Regarding the independence of the nominator and denominator of the SINR expression (2), this quantity can be further reduced to

$$\mathbb{E}\gamma_{u_cc,n} = \left(\rho_{u_cc} P_{c,n} \mathbb{E}\|\boldsymbol{h}_{u_cc,n}\|^2 |\tilde{\boldsymbol{h}}_{u_cc,n}^\dagger \hat{\boldsymbol{f}}_{c,n}|^2\right) \mathbb{E}\left(\sigma^2 N_t + \sum_{c' \in \mathcal{C}_n^a / \{c\}} \tilde{\rho}_{u_cc'} P_{c',n} \|\boldsymbol{g}_{u_cc',n}\|^2 |\tilde{\boldsymbol{g}}_{u_cc',n}^\dagger \hat{\boldsymbol{f}}_{c',n}|^2\right)^{-1}. \tag{33}$$

Consider $\mathbb{E}\|\boldsymbol{h}_{u_cc,n}\|^2 |\tilde{\boldsymbol{h}}_{u_cc,n}^\dagger \hat{\boldsymbol{f}}_{c,n}|^2$. Following the same lines presented in Appendix A, we have

$$|\tilde{\boldsymbol{h}}_{u_cc,n}^\dagger \hat{\boldsymbol{f}}_{c,n}|^2 = \cos^2(\theta_{u_cc,n})|\hat{\boldsymbol{h}}_{u_cc,n}^\dagger \hat{\boldsymbol{f}}_{c,n}|^2 + \sin^2(\theta_{u_cc,n})|\boldsymbol{s}_{u_cc,n}^\dagger \hat{\boldsymbol{f}}_{c,n}|^2 + \sin(2\theta_{u_cc,n})\Re(\hat{\boldsymbol{h}}_{u_cc,n}^\dagger \hat{\boldsymbol{f}}_{c,n} \hat{\boldsymbol{f}}_{c,n}^\dagger \boldsymbol{s}_{u_cc,n}), \tag{34}$$

where $\Re(.)$ indicates the real part. Since for any two vectors $\boldsymbol{x}$ and $\boldsymbol{y}$ there holds $\Re(\boldsymbol{x}^\dagger \boldsymbol{y}) \leq |\boldsymbol{x}||\boldsymbol{y}|$, we have $\Re(\hat{\boldsymbol{h}}_{u_cc,n}^\dagger \hat{\boldsymbol{f}}_{c,n} \hat{\boldsymbol{f}}_{c,n}^\dagger \boldsymbol{s}_{u_cc,n}) \leq \left|\hat{\boldsymbol{h}}_{u_cc,n}^\dagger \hat{\boldsymbol{f}}_{c,n}\right| \left|\boldsymbol{s}_{u_cc,n}^\dagger \hat{\boldsymbol{f}}_{c,n}\right| \leq 1$ Thus, $|\tilde{\boldsymbol{h}}_{u_cc,n}^\dagger \hat{\boldsymbol{f}}_{c,n}|^2 \leq \cos^2(\theta_{u_cc,n})|\hat{\boldsymbol{h}}_{u_cc,n}^\dagger \hat{\boldsymbol{f}}_{c,n}|^2 + \sin^2(\theta_{u_cc,n})|\boldsymbol{s}_{u_cc,n}^\dagger \hat{\boldsymbol{f}}_{c,n}|^2 + \sin(2\theta_{u_cc,n})$. We then have

$$\mathbb{E}\|\boldsymbol{h}_{u_cc,n}\|^2 |\tilde{\boldsymbol{h}}_{u_cc,n}^\dagger \hat{\boldsymbol{f}}_{c,n}|^2 \approx (N_t - |\mathcal{C}|)\left(1 - \frac{N_t - 1}{N_t}\delta_{u_cc,n}\right) + \delta_{u_cc,n} + N_t \mathbb{E}\sin(2\theta_{u_cc,n}) = \hat{\delta}_{u_cc,n}, \tag{35}$$

Evaluation of $\mathbb{E}\sin(2\theta_{u_cc,n})$ is tricky as we only know the pdf of random variable $\sin^2(\theta_{u_cc,n})$. Consider the upper-bound $\mathbb{E}\sin(2\theta_{u_cc,n}) \leq \mathbb{E}|\sin(2\theta_{u_cc,n})|$, which is equal to $\mathbb{E}\sqrt{\sin^2(2\theta_{u_cc,n})}$. Note that the inequality is tight when $0 \leq \theta_{u_cc,n} \leq \pi/2$. A straightforward manipulation suggests that the pdf of random variable $\sqrt{\sin^2(2\theta_{u_cc,n})}$ is

$$f_{\sqrt{\sin^2(2\theta_{u_cc,n})}}(x) = \frac{x}{2\sqrt{1-x^2}}\left(f_{\sin^2(\theta_{u_cc,n})}\left(\frac{1+\sqrt{1-x^2}}{2}\right) + f_{\sin^2(\theta_{u_cc,n})}\left(\frac{1-\sqrt{1-x^2}}{2}\right)\right), \tag{36}$$

where $0 \leq x \leq 1$. Using the derived pdf, it is straightforward to confirm that

$$\mathbb{E}\sin(2\theta_{u_cc,n}) \leq \int_0^{\delta_{u_cc,n}} \frac{x^2}{2\sqrt{1-x^2}} \mathrm{f}_{\sin^2(\theta_{u_cc,n})}\left(\frac{1+\sqrt{1-x^2}}{2}\right) dx + \int_0^{\delta_{u_cc,n}} \frac{x^2}{2\sqrt{1-x^2}} \mathrm{f}_{\sin^2(\theta_{u_cc,n})}\left(\frac{1-\sqrt{1-x^2}}{2}\right) dx. \quad (37)$$

We then only need to substitute (37) back into (35) to obtain an approximate of $\hat{\delta}_{u_cc,n}$. Now, consider the term $\mathbb{E}(\sigma^2 N_t + \sum_{c' \in \mathcal{C}_n^u/\{c\}} \tilde{\rho}_{u_cc'} P_{c',n} J_{u_cc',n})^{-1}$ in (33), which can be evaluated as

$$\mathbb{E}\int_0^\infty e^{-w\left(\sigma^2 N_t + \sum_{c' \in \mathcal{C}_n^u/\{c\}} \tilde{\rho}_{u_cc'} P_{c',n} J_{u_cc',n}\right)} dw = \int_0^\infty e^{-w\sigma^2 N_t} \prod_{c' \in \mathcal{C}_n^u/\{c\}} \frac{1}{1 + \tilde{\rho}_{u_c'} P_{c',n}\delta_{u_cc'} w} dw. \quad (38)$$

Substituting (38) into (33), the desired result is proved.

## C. Proof of Result 3

According (35), we approximate the received signal strength by

$$\rho_{u_cc} P_{c,n} \|\boldsymbol{h}_{u_cc,n}\|^2 |\tilde{\boldsymbol{h}}_{u_cc,n}^\dagger \hat{\boldsymbol{f}}_{c,n}|^2 \approx \rho_{u_cc} P_{c,n} \left(1 - \frac{N_t - 1}{N_t}\delta_{u_cc,n}\right) \chi^2_{2(N_t - |\mathcal{C}|)}. \quad (39)$$

We define another random variable $Y = \sum_{c' \neq c} \tilde{\rho}_{u_cc'} P_{c',n} J_{u_cc',n}$. We introduce a new random variable defined as $Z = \frac{\rho_{u_cc} P_{c,n}\left(1 - \frac{N_t - 1}{N_t}\delta_{u_cc,n}\right)\chi^2_{2(N_t - |\mathcal{C}|)}}{\sigma^2 N_t + Y}$. . We start by deriving an expression for the pdf of random variable $Z$ as follows:

$$\mathrm{f}_Z(z) = \frac{\partial}{\partial z}\mathbb{P}\left\{\frac{\chi^2_{2(N_t - |\mathcal{C}|)}}{\sigma^2 N_t + Y} \leq \frac{z}{\rho_{u_cc} P_{c,n}\left(1 - \frac{N_t - 1}{N_t}\delta_{u_cc,n}\right)}\right\}$$

$$= \mathbb{E}_Y\left[\frac{\sigma^2 N_t + Y}{\rho_{u_cc} P_{c,n}\left(1 - \frac{N_t - 1}{N_t}\delta_{u_cc,n}\right)}\mathrm{f}_{\chi^2_{2(N_t - |\mathcal{C}|)}}\left(z\frac{\sigma^2 N_t + Y}{\rho_{u_cc} P_{c,n}\left(1 - \frac{N_t - 1}{N_t}\delta_{u_cc,n}\right)}\right)\right]$$

$$= \mathbb{E}_Y\left[\left(\frac{\sigma^2 N_t + Y}{\rho_{u_cc} P_{c,n}\left(1 - \frac{N_t - 1}{N_t}\delta_{u_cc,n}\right)}\right)^{N_t - |\mathcal{C}|}\frac{z^{N_t - |\mathcal{C}| - 1}e^{-z\frac{\sigma^2 N_t + Y}{\rho_{u_cc} P_{c,n}\left(1 - \frac{N_t - 1}{N_t}\delta_{u_cc,n}\right)}}}{\Gamma(N_t - |\mathcal{C}|)}\right]$$

$$= \frac{z^{N_t - |\mathcal{C}| - 1}}{\Gamma(N_t - |\mathcal{C}|)}\mathbb{E}_Y\left[(-1)^{N_t - |\mathcal{C}|}\frac{\partial^{N_t - |\mathcal{C}|}}{\partial z^{N_t - |\mathcal{C}|}}e^{-z\frac{\sigma^2 N_t + Y}{\rho_{u_cc} P_{c,n}\left(1 - \frac{N_t - 1}{N_t}\delta_{u_cc,n}\right)}}\right],$$

where we substituted the pdf of random variable $\chi^2_{2(N_t - |\mathcal{C}|)}$ and applied some straightforward manipulations. Consequently,

$$\mathrm{f}_Z(z) = \frac{(-1)^{N_t - |\mathcal{C}| + 1}z^{N_t - |\mathcal{C}|}}{\Gamma(N_t - |\mathcal{C}|)}\frac{\partial^{N_t - |\mathcal{C}|}}{\partial z^{N_t - |\mathcal{C}|}}\mathbb{E}_Y\left[e^{-z\frac{\sigma^2 N_t + Y}{\rho_{u_cc} P_{c,n}\left(1 - \frac{N_t - 1}{N_t}\delta_{u_cc,n}\right)}}\right],$$

which, by recalling the definition of random variable $Y$ and after some straightforward manipulation, can be written as

$$\mathrm{f}_Z(z) = (-1)^{N_t - |\mathcal{C}|}\frac{z^{N_t - |\mathcal{C}| - 1}}{\Gamma(N_t - |\mathcal{C}|)}\frac{\partial^{N_t - |\mathcal{C}|}O(z)}{\partial z^{N_t - |\mathcal{C}|}}, \quad (40)$$

where $O(z)$ is defined as

$$O(z) = e^{-z\frac{\sigma^2 N_t}{\rho_{u_c c}P_{c,n}\left(1-\frac{N_t-1}{N_t}\delta_{u_c c,n}\right)}} \prod_{c' \in \mathcal{C}_n^a/\{c\}} \frac{1}{1+z\frac{\tilde{\rho}_{u_c c'}P_{c',n}\delta_{u_c c'}}{\rho_{u_c c}P_{c,n}\left(1-\frac{N_t-1}{N_t}\delta_{u_c c,n}\right)}}. \tag{41}$$

Utilizing pdf (40), get the following expression:

$$\mathbb{E}\left[e^{-\theta_{u_c}\log(1+Z)}\right] = \frac{(-1)^{N_t-|\mathcal{C}|}}{\Gamma(N_t-|\mathcal{C}|)} \int_0^\infty \frac{z^{N_t-|\mathcal{C}|-1}}{(1+z)^{\theta_{u_c}}}\frac{\partial^{N_t-|\mathcal{C}|}O(z)}{\partial z^{N_t-|\mathcal{C}|}}dz. \tag{42}$$

Using (42), we derive an approximate of the effective capacity as suggested in Result 3.

### D. Proof of Result 4

It is straightforward to show that

$$\mathbb{E}e^{-\theta_{u_c}\log(1+\gamma_{u_c c,n})} = 1 - \theta_{u_c}\bar{R}_{u_c c,n} + \sum_{k=2}^\infty \frac{(-\theta_{u_c})^k}{k!}\mathbb{E}\left(\log(1+\gamma_{u_c c,n})\right)^k, \tag{43}$$

where $\bar{R}_{u_c c,n}$ is already known from Result 1 or Result 2. Recalling the definitions of the random variables $J_{u_c c,n}$ and $K_{u_c c,n}$ from Appendix A, we get

$$\mathbb{E}\left(\log(1+\gamma_{u_c c,n})\right)^k = \mathbb{E}\left(\int_0^\infty \frac{e^{-w}}{w}(1-e^{-w\gamma_{u_c c,n}})dw\right)^k = \mathbb{E}\int_0^\infty \cdots \int_0^\infty \prod_{i=1}^k \frac{e^{-w_i}}{w_i}(1-e^{-w_i\gamma_{u_c c,n}})dw_i$$

$$= \int_0^\infty \cdots \int_0^\infty \left(\prod_{i=1}^k \frac{e^{-\sigma^2 N_t w_i}}{w_i}\right)\mathbb{E}e^{-\tilde{\rho}_{u_c c'}P_{c',n}J_{u_c c,n}\sum_{i=1}^k w_i}\mathbb{E}\prod_{i=1}^k\left(1-e^{-\rho_{u_c c}P_{c,n}K_{u_c c,n}w_i}\right)dw_1 \ldots dw_k$$

$$= \int_0^\infty \cdots \int_0^\infty \frac{e^{-\sigma^2 N_t \sum_{i=1}^k w_i}}{\prod_{i=1}^k w_i}\prod_{c' \in \mathcal{C}_n^a/\{c\}}\frac{1}{1+\tilde{\rho}_{u_c c'}P_{c',n}\delta_{u_c c',n}\sum_{i=1}^k w_i}\mathbb{E}\prod_{i=1}^k\left(1-e^{-\rho_{u_c c}P_{c,n}K_{u_c c,n}w_i}\right)dw_1 \ldots dw_k$$

For the general case, it is too complicated to derive a closed-form expression for this integral. We therefore assume $k=2$, and denote $\hat{R}_{u_c c_n} = \mathbb{E}\left(\log(1+\gamma_{u_c c,n})\right)^2$. As a result,

$$\hat{R}_{u_c c_n} = \int_0^\infty \int_0^\infty \frac{e^{-\sigma^2 N_t(w_1+w_2)}}{w_1 w_2}\prod_{c' \in \mathcal{C}_n^a/\{c\}}\frac{1}{1+\tilde{\rho}_{u_c c'}P_{c',n}\delta_{u_c c',n}(w_1+w_2)}(1-\hat{K}(w_1,w_2))dw_1 dw_2. \tag{44}$$

where, by following the lines presented in the proof of Result 1, it is straightforward to confirm that $\hat{K}(w_1, w_2)$ is obtained as

$$\hat{K}(w_1, w_2) = \int_0^{\delta_{u_c c,n}} f_{\sin^2(\theta_{u_c c,n})}(x)\left[\frac{\frac{1}{1+w_1 P_{c,n}\rho_{uc}\delta_{u_c c,n}}}{(1+w_1 P_{c,n}\rho_{uc}(1-x))^{N_t-|\mathcal{C}_n^a|}}\right.$$

$$\left. +\frac{\frac{1}{1+w_2 P_{c,n}\rho_{uc}\delta_{u_c c,n}}}{(1+w_2 P_{c,n}\rho_{uc}(1-x))^{N_t-|\mathcal{C}_n^a|}}-\frac{\frac{1}{1+(w_1+w_2)P_{c,n}\rho_{uc}\delta_{u_c c,n}}}{(1+(w_1+w_2)P_{c,n}\rho_{uc}(1-x))^{N_t-|\mathcal{C}_n^a|}}\right]dx.$$

Substituting (44) into (43), the proposed approximation in Result 4 is then obtained.